\documentclass[twocolumn]{aastex63}

\usepackage{multirow}

\received{December 2021}

\shorttitle{How does environment affect the morphology of radio AGN?}
\shortauthors{Morris et al.}

\graphicspath{{./}{figures/}}

\begin{document}

\title{How does environment affect the morphology of radio AGN?}

\author[0000-0001-9920-0210]{Melissa Elizabeth Morris}
\affiliation{UW Madison}
\author{Eric Wilcots}
\affiliation{UW Madison}
\author{Eric Hooper}
\affiliation{UW Madison}
\author{Sebastian Heinz}
\affiliation{UW Madison}

\begin{abstract}

Galaxies hosting Active Galactic Nuclei (AGN) with bent radio jets are used as tracers of dense environments, such as galaxy groups and clusters. The assumption behind using these jets is that they are bent under ram pressure from a dense, gaseous medium through which the host galaxy moves. However, there are many AGN in groups and clusters with jets that are not bent, which leads us to ask: why are some AGN jets affected so much by their environment while others are seemingly not? We present the results of an environmental study on a sample of 185 AGN with bent jets and 191 AGN with unbent jets in which we characterize their environments by searching for neighboring galaxies using a Friends-of-Friends algorithm. We find that AGN with bent jets are indeed more likely to reside in groups and clusters, while unbent AGN are more likely to exist in singles or pairs. When considering only AGN in groups of 3 or more galaxies, we find that bent AGN are more likely to exist in halos with more galaxies than unbent AGN. We also find that unbent AGN are more likely than bent AGN to be the brightest group galaxy. Additionally, groups hosting AGN with bent jets have a higher density of galaxies than groups hosting unbent AGN. Curiously, there is a population of AGN with bent jets that are in seemingly less dense regions of space, indicating they may be embedded in a cosmic web filament. Overall, our results indicate that bent doubles are more likely to exist in in larger, denser, and less relaxed environments than unbent doubles, potentially linking a galaxy's radio morphology to its environment.

\end{abstract}


\section{Introduction} \label{sec:intro}

Radio galaxies have very distinct morphologies, extending far from the optical host galaxy in two straight, biconical jets pointed in opposite directions. However, there are jets whose morphologies deviate from this norm, appearing bent instead of straight. The general consensus is that these jets are likely bent by ram pressure from the gaseous medium in which they are embedded \citep{burnsowen1980,jonesowen1979,blanton2000,freeland2008,freeland2011,wingblanton2011,morsony2013}. This hypothesis is supported by the fact that these radio galaxies are prevalent in dense environments, such as galaxy clusters \citep{owenrudnick1976,blanton2000,blanton2001,silverstein2018,mingo2019}. Indeed, they have been used to identify galaxy clusters at higher redshifts \citep{blanton2015,goldenmarx2021}.

These objects have had a variety of names, such as head-tail radio sources, Wide Angle Tail (WAT)/Narrow Angle Tail (NAT) sources, or bent double lobed sources \citep{owenrudnick1976,ekers1978,begelman1979,blanton2000,wingblanton2011}. The WAT/NAT designation is one that depends on the degree of bending, with NATs being extremely bent and WATs being less bent. Several studies have found that WATs are likely to be the brightest cluster/group galaxy, indicating that they are at or near the center of their larger host halo \citep{sakelliou2000,blanton2001}. However, many more recent studies have found that they exist in a wide array of environments, many of them existing well beyond the virial radius of the nearest cluster \citep{garon2019,devos2021,goldenmarx2021}.

Studies have found AGN with bent jets in less dense environments, as well. There have been many bent AGN, especially WATs that are not as bent as NATS, found in galaxy groups \citep{ekers1978,freeland2008,freeland2011}. \cite{edwards2010} discovered a bent AGN in a filament, and other studies have found populations of bent doubles in pairs or isolated environments that are less likely to have an intergalactic medium (IGM) dense enough to exert ram pressure to bend the jets \citep{blanton2001,garon2019}.

Unbent double lobed radio sources are also found in dense environments \citep{worpel2013,ching2017,garon2019}, yet they are not affected by the surrounding medium in the same way that bent double lobed AGN are. This could be indicative of a difference in the gas density and dynamics of the environments of these two kinds of galaxies. Additionally, it could indicate that bent and unbent AGN exist in different types of cluster or group galaxies -- while unbent AGN typically exist in the brightest group or cluster galaxies \citep{pasini2021}, bent AGN may exist in other galaxies that are moving through their environment at higher velocities, allowing for their jets to be bent by ram pressure. Alternatively, the medium in environments where bent jets are found could be undergoing bulk motion, causing the jets to appear bent. Comparing environments of bent and unbent AGN could help us understand the mechanisms responsible for bending of AGN jets. \cite{wingblanton2011} found that bent AGN were more often found in clusters and denser environments than unbent galaxies. However, it is still unclear what further differences exist between bent and unbent AGN in dense environments and why some AGN in dense environments are bent while others are not.

There exist large samples of radio galaxies that have been morphologically classified. For example, the Radio sources associated with Optical Galaxies and having Unresolved or Extended morphologies (ROGUE) catalog \citep{kozielwierzbowska2020} uses observations from a variety of radio and optical surveys to create a catalog of radio sources of various morphologies. Additionally, large samples of bent doubles have been identified using image recognition techniques \citep{proctor2006} on data from the Faint Images of the Radio Sky at Twenty-cm \citep[FIRST;][]{becker1995} survey, but still little is known about their environment. However, deep photometric surveys like the Dark Energy Camera Legacy Survey \citep[DECaLS;][]{dey2019} and the unblurred coadds of Wide-field Infrared Survey Explorer images \citep[unWISE;][]{lang2014}, which reach galaxies with absolute magnitudes r$<$23, reveal many galaxies that could potentially neighbor the AGN hosts.

Throughout the literature, Friends of Friends (FoF) algorithm have been used to identify galaxy groups and to better understand the environments of galaxies \citep{huchrageller1982,tempel2014,duartemamon2014}. Most of these studies rely on spectroscopic redshifts due to their low uncertainties and high precision. However, spectroscopic redshifts are observationally expensive. Few studies have attempted to adapt FoF algorithms to use photometric redshifts \citep{liu2008,jian2014}, which are much less costly observationally but have higher uncertainties that can potentially lead to contamination in the group finding results. Nonetheless, a FoF algorithm that relies on photometric redshifts can still prove useful in understanding the environments of bent doubles and comparing them to those of unbent doubles.

The goal of this study is to use an FoF algorithm combined with photometric redshifts from unWISE and DECaLS \citep{lang2014,dey2019,zou2019} to identify galaxies neighboring bent and unbent doubles, and then to use these galaxies to classify the environments of these two groups of AGN. This will aid in the understanding of how the environment affects the morphology of radio AGN, and give us a better idea of what kind of environments are being probed by bent double lobed AGN as opposed to unbent AGN. In Section 2, we outline the radio and optical data used in this study and how we selected our bent and unbent samples. In Section 3, we go into detail about how our FoF algorithm works and how it incorporates photometric redshifts to find neighbors, as well as how we calibrated parameters for the FoF algorithm and tested it using the cosmological simulation Illustris TNG. In Section 4, we outline the broad results of our FoF algorithm on the data by taking measurements of the frequency of AGN in different environments, galaxy density of these environments, and r-band magnitude gaps. In Section 5, we discuss these results and in Section 6, we outline our main conclusions for this study. Throughout this paper, we use cosmological parameters from the WMAP9 results \citep[][]{hinshaw2013}, H$_0$ = 69.3 km s$^{-1}$ Mpc$^{-1}$ and $\Omega_0$ = 0.287.

\begin{figure*}
    \centering
    \includegraphics[scale=.7]{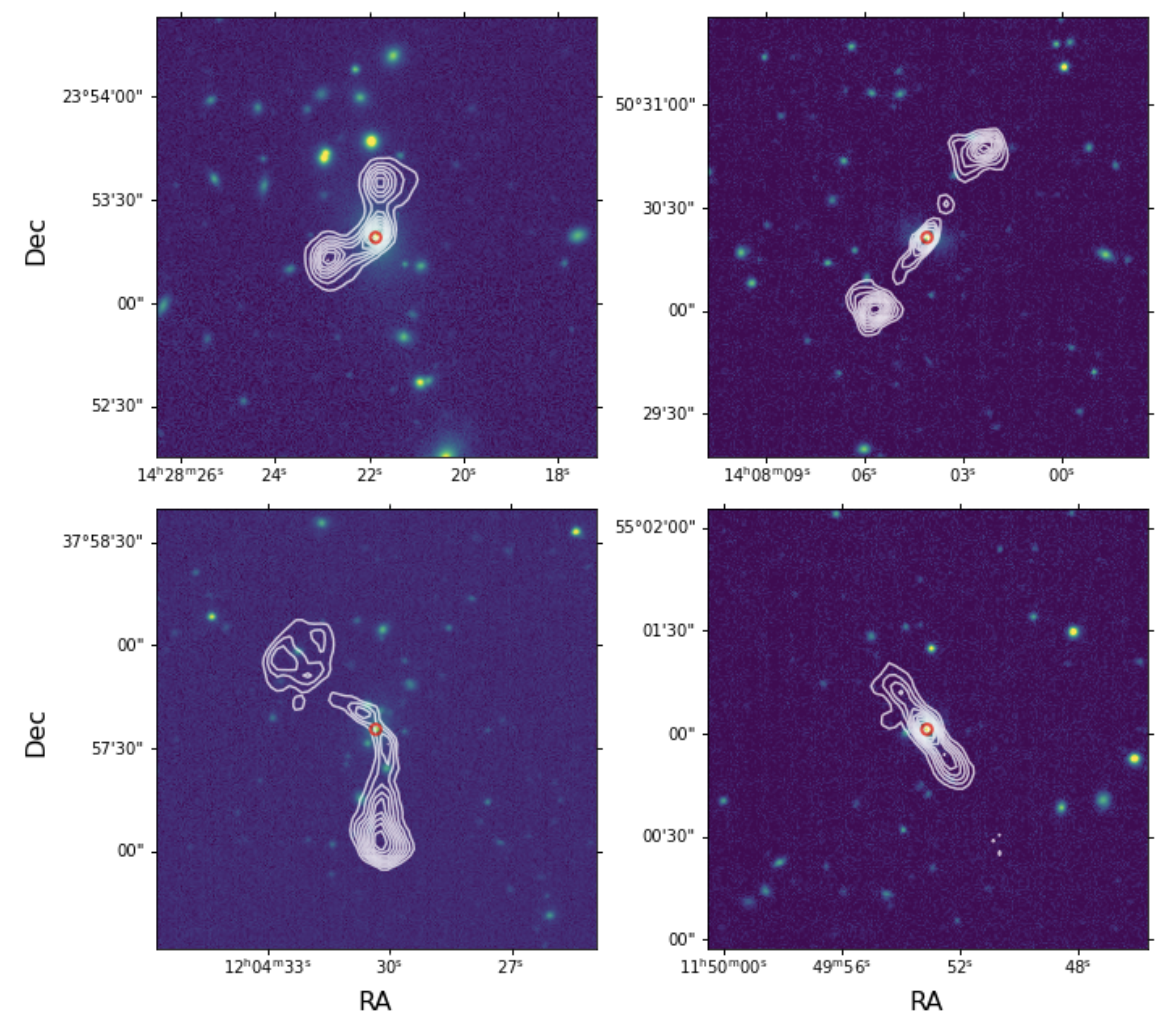}
    \caption{3$\sigma$ contours from FIRST plotted over DECaLS r-band images, all of which are 2 by 2 arcminutes on each side. The red circle indicates the position of the host galaxy in SDSS. Left side: 2 examples of bent doubles. Right side: 2 examples of unbent AGN.}
    \label{fig:vis_select}
\end{figure*}

\begin{figure}
    \centering
    \includegraphics[scale=.6]{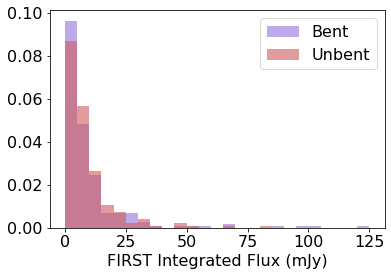}
    \caption{The distribution of FIRST integrated fluxes for the bent and unbent samples.}
    \label{fig:firstflux}
\end{figure}

\section{Data} \label{sec:data}

We carried out this study using a sample of 185 visually confirmed bent double lobed AGN and 191 visually confirmed FRI and FRII AGN that are not bent. Of these samples, all of the unbent AGN and 83 bent AGN were identified in the ROGUE catalog \citep{kozielwierzbowska2020}, which combines FIRST \citep{becker1995}, NRAO VLA Sky Survey \citep[NVSS;][]{condon1998}, and Sloan Digital Sky Survey \citep[SDSS;][]{abazajian2009} spectroscopy to identify radio sources and categorize them morphologically. Examples from the final sample of bent and unbent doubles can be seen in Figure \ref{fig:vis_select} and the distributions of integrated flux densities measured by FIRST are in Figure \ref{fig:firstflux}.

To better understand what kinds of environments bent double lobed AGN occupy, we use an FoF algorithm combined with photometric redshifts \citep{zou2019} derived from two deep all-sky surveys, the DECaLS \citep{dey2019} and unWISE, \citep{lang2014} which reach magnitudes of r $<$ 23. To determine halo membership in redshift-space, we use photometric redshifts and stellar masses derived from the g, r, and z bands from DECaLS and the W1 and W2 bands from un-WISE \citep{zou2019}.

\subsection{Bent and Unbent Double Selection}

Of the sample of 185 bent doubles used for this study, 102 were chosen from a larger sample of 882 galaxies identified in the FIRST survey by the automated image recognition techniques of \cite{proctor2006}, which categorized radio galaxies as bent based on their morphology in FIRST. From this sample, 261 sources had optical counterparts that overlapped with the FIRST source whose spectroscopic redshifts were measured in SDSS DR7. We select these galaxies because we know their location in redshift space very precisely, so that when we perform our FoF algorithm, we know exactly what redshift we should be comparing to.

From this sample, we visually identified bent doubles based on the radio morphology with respect to the host galaxy. Examples of galaxies that have been visually selected are shown in the left two panels of Figure \ref{fig:vis_select}. The radio contours are in white and the location of the host galaxy is circled in red. This visual identification left us with an initial sample of 160 visually confirmed bent doubles with measured spectroscopic redshifts. The 101 radio galaxies identified by \cite{proctor2006} that we left out of our sample either did not visually resemble bent doubles or were not associated with the SDSS source. While it can be difficult to determine the precise morphology for many of our bent sources due to resolution limitations, we can say that at least 75$\%$ of the bent double sources in our sample are WATs, which is slightly higher than \cite{bhukta2021} (71$\%$ WATs) and \cite{pal2021} (65$\%$ WATs) and significantly higher than \cite{mingo2019} (42$\%$ WATs).

For unbent AGN and the rest of our bent AGN sample, we used the ROGUE catalog \citep{kozielwierzbowska2020}, which morphologically classified radio counterparts in FIRST and NVSS for galaxies with observed spectra in SDSS DR7. From their sample, we looked only at FRI and FRII galaxies and selected 191 unbent AGN based on a visual confirmation of the radio contours in FIRST with respect to the optical host galaxy. Of these, 64 are FRI radio galaxies and 127 are FRII radio galaxies. The right images in Figure \ref{fig:vis_select} are two examples from this sample. We also selected an additional 83 bent doubles by looking at galaxies they identified as WATs or NATs. These galaxies did not appear in the selection from \cite{proctor2006} possibly due to the fact that the ROGUE catalog relies not only on FIRST, but also on NVSS, to identify AGN. It is possible that, without these additional observations, \cite{proctor2006} was unable to identify these sources as bent AGN.

\begin{figure}
    \centering
    \includegraphics[scale=.6]{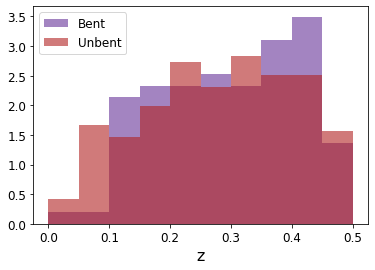}
    \caption{Normalized histograms comparing the redshift distributions of bent and unbent AGN in our sample.}
    \label{fig:zdist}
\end{figure}

While both samples were visually classified, they were selected using slightly different radio criteria. The bent sample from \cite{proctor2006} only used FIRST contours, while galaxies selected from the ROGUE catalog used both FIRST and NVSS, which resulted in a slightly lower redshift distribution. In order to avoid redshift-dependent differences between the two samples, we limit the rest of this analysis to only those galaxies below a redshift of 0.5, beyond which there are no unbent AGN in our sample. This reduced our sample of 243 bent AGN to a total of 185 bent doubles. The FIRST Flux distributions can be found in Figure \ref{fig:firstflux} and shows that these samples are similar in their fluxes. The redshift distributions of both samples can be found in Figure \ref{fig:zdist} and are fairly consistent with each other to minimize redshift-dependent differences between the two populations.

\subsection{DECaLS and unWISE}
\begin{figure}
    \centering
    \includegraphics[scale=.4]{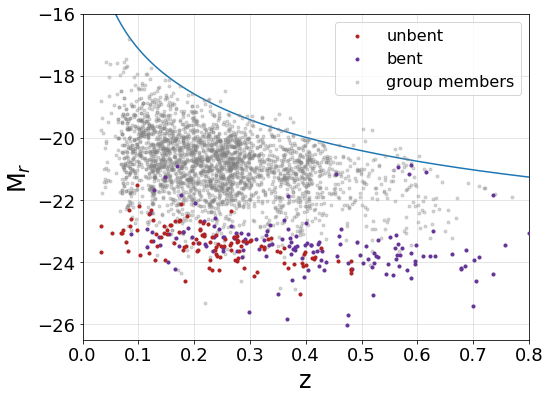}
    \caption{Absolute r-band magnitude as a function of photometric redshift for the bent and unbent samples, as well as all identified group members. The blue line signifies an r=21 limiting apparent magnitude. The majority of radio galaxies are below an absolute magnitude of -22, making up the brightest galaxies. It is important to note that only galaxies below z = 0.5 are considered in this study.}
    \label{fig:maglim}
\end{figure}

DECaLS and unWISE are two surveys that provide deep (r $<$ 23) photometric data across the sky in the g, r, and z bands for DECaLS and the W1 and W2 filters for unWISE \citep{dey2019}. To determine halo membership, we use the work of \cite{zou2019}, which measures the k-corrected absolute magnitudes, stellar mass, and photometric redshifts for galaxies observed by DECaLS and unWISE.

\begin{figure}
    \centering
    \includegraphics[scale=.7]{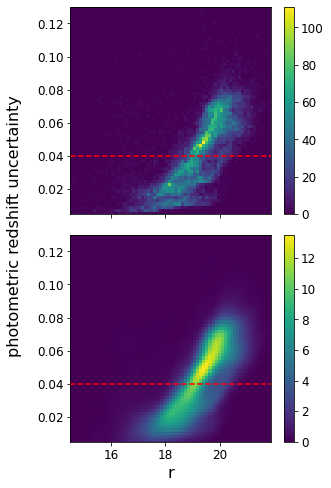}
    \caption{(Top) A 2D histogram from DECaLS galaxies of the photometric redshift uncertainty as a function of apparent r-band magnitude. (Bottom) A Gaussian Kernel Density Estimate (KDE) of the top plot. In both plots, the red dashed line represents an uncertainty of 0.04, which is what we have chosen as our redshift uncertainty cutoff. This results in many fainter (r$>$20) galaxies being excluded from this analysis, which can be seen in Figure \ref{fig:maglim}}.
    \label{fig:2dhists}
\end{figure}

Figure \ref{fig:maglim} plots the k-corrected absolute magnitudes of the unbent and bent double sample, as well as all of the group members identified by the FoF algorithm. For reference the blue line represents an apparent magnitude of r $=$ 21. While DECaLS and unWISE do reach r $<$ 23, we find that the galaxies with low photometric errors tend to be the brighter galaxies, as can be seen in Figure \ref{fig:2dhists}. Therefore, when applying a photometric error cutoff, we tend to exclude many of the dimmer galaxies that fall between 21 $<$ r $<$ 23. Also of note is that the radio galaxies are on the whole much brighter than the other galaxies occupying the same halos.

\section{Developing and Testing the Group Finding Algorithm} \label{sec:FoF}
\subsection{Defining the Algorithm}
We use a Friends of Friends algorithm to identify possible galactic neighbors using their positions in the sky and measured photometric redshifts from DECaLS and unWISE \citep{zou2019}. This method has been used throughout the literature to identify potential galaxy groups, though it is usually only applied to galaxies with measured spectroscopic redshifts \citep{tempel2014,duartemamon2014}. Errors on photometric redshifts are typically larger than potential group or even cluster velocity dispersions, making them particularly susceptible to false positives (i.e. falsely identifying a galaxy as a group member). In order to minimize the likelihood of false positives, we use a probability motivated approach when considering the locations of galaxies in redshift-space, while using the FoF method when considering galaxy positions in the plane of the sky.

We calibrated and tested the algorithm on mock images created using the Illustris TNG300 simulation, which simulated a box of cosmic space that is 300 Mpc long on each side. This allowed us to sample a wide variety of different environments and see how the FoF algorithm performed on each of them.

\subsubsection{FoF: Plane of the sky}
A typical FoF algorithm starts at the position of an initial galaxy and searches within a predetermined radius, known as the linking length, for nearby neighboring galaxies. When a neighboring galaxy is found within one linking length of the original galaxy, another search is done for additional galaxies within one linking length of the neighboring galaxy. This process is repeated for all galaxies that fall within one linking length of another galaxy until no more have been detected. Our FoF algorithm starts at the bent double and only includes galaxies that fulfill the redshift criteria outlined in Section \ref{sec:prob}.

\begin{figure}
    \centering
    \includegraphics[scale=0.6]{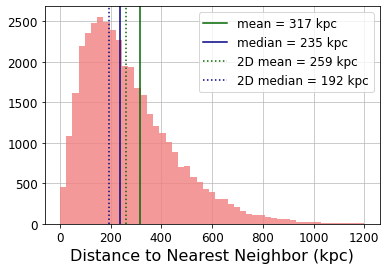}
    \caption{The distribution of 3D distances to the nearest neighbor for all galaxies in groups of at least 5 members in the Illustris TNG300 simulations. This histogram only takes into account galaxies that would be observable by DECaLS at a redshift of roughly z = 0.1. The vertical lines show different characteristics of this distribution which were tested as linking lengths for our FoF algorithm.}
    \label{fig:nearneigh}
\end{figure}

There are a variety of linking lengths in the literature \citep{duartemamon2014}, but many of these are established using data sets that vary drastically from ours in their completeness, largely due to the fact that they rely on spectroscopic observations instead of photometric observations. Spectroscopic observations allow them to scale the linking length by the number density of galaxies, calculated precisely due to the fact that spectroscopic redshifts have much lower uncertainties than photometric redshifts. Because we are relying on photometric redshifts, such scaling becomes unreliable. Therefore, we have chosen a static value of 300 kpc, which is motivated by the distribution of the distance to the nearest group galaxy as given by Illustris (see Section \ref{sec:caliblinlen}). This linking length will include the majority of the galaxies in the same group as the bent double based on the plot in Figure \ref{fig:nearneigh}. We also set a maximum distance from the bent double at 2 Mpc, as groups are unlikely to extend that far and if the algorithm does attempt to go that far, it is likely due to a false positive detection \citep{tempel2014,silverstein2018}.

\subsubsection{Probability based: Redshift space} \label{sec:prob}
Because typical photometric redshift uncertainties tend to be much larger than expected redshift variations due to galaxy group/cluster velocity dispersions, we could not apply the FoF algorithm in redshift space. Instead, we devised a method of measuring the relative probability that two galaxies are at the same redshift based upon their uncertainties and set a cutoff, where any galaxies that fell below this were not considered to be at the redshift of the bent double.

We first assumed that the probability density functions $P$ could be described by normal distributions with the mean $\mu$ being the measured redshift and the standard deviation $\sigma$ being the uncertainties on said redshift. By multiplying the probability density functions for the bent double and the potential neighbor and normalizing by the maximum probability, we found the probability that these two galaxies reside at the same redshift.

The maximum possible probability was calculated by multiplying the probability distributions, assuming they had the same redshift $\mu_o$ with their measured uncertainties $\sigma_1$ and $\sigma_2$. The choice of $\mu_o$ is less relevant than the choices of the uncertainties.

$$P_{\text{max}} = \int P(\mu_o,\sigma_1) P(\mu_o,\sigma_2)$$

The final probability was taken by multiplying the real probability density functions using $\mu$ and $\sigma$ as the measured redshift and redshift uncertainty and dividing by $P_{\text{max}}$:

$$P = \frac{\int P(\mu_1,\sigma_1) P(\mu_2, \sigma_2)}{P_{\text{max}}}$$

Spectroscopic measurements of the bent double were always used for $P(\mu_1,\sigma_1)$ since these redshifts were all more accurate than photometric redshifts. Therefore, the redshifts of potential neighbor galaxies were always compared to the redshift of the bent double, as opposed to being compared to one another. If the final probability that the potential neighbor was at the redshift of the bent double was less than 0.2 (motivated in Section \ref{sec:pickprob}), it was removed from the group to reduce the number of potential false positives as much as possible.

\subsection{Creating Mock Catalogs from Illustris TNG}

To understand how well our algorithm performs and tune various parameters, we carried out a series of tests using mock data created from the Illustris TNG300 simulations, which evolves a cubic volume with 300 Mpc on every side \citep{nelson2019}. We used catalogs of subhalos (potential galaxy analogs) and groups of dark matter subhalos produced by the Illustris team as our starting point for generating mock observational data. We then ran the FoF algorithm described above to tune the parameters and find the best balance between high percent completeness for each individual group and low percent of false group member identifications.

We started this analysis by creating a mock catalog of galaxy positions and assigning photometric redshift measurements based on their positions in the simulation volume. Using the subhalo catalog from the z = 0.10 snapshot of the TNG300 simulation volume, we narrowed our search by selecting only subhalos with absolute r-band magnitudes $< -16$ in order to immediately eliminate any subhalos that would not be visible with DECaLS. The entire simulation volume was then projected along the z-axis such that the front end was assigned z = 0.10 and redshifts were calculated for each subhalo using their position in the simulation volume along the z-axis. From here, right ascensions and declinations were calculated for each galaxy based on their position in the xy plane of the simulation and their calculated redshift.

Our next goal was to make this idealized catalog seem more like real DECaLS data. We started by calculating apparent r-band magnitudes for each subhalo and then assigning a redshift uncertainty based on real uncertainties from DECaLS, which vary with r, as shown in Figure \ref{fig:2dhists}. Using a 2D histogram of DECaLS photometric redshift uncertainties and r-band magnitudes, we applied a Gaussian kernel density estimate to smooth the histogram (as seen in Figure \ref{fig:2dhists}) and then, for each simulation subhalo, sampled from a range of uncertainties based on their calculated apparent r-band magnitudes. We then added a small offset to the redshift, based on the randomly assigned uncertainty. After this, we applied an apparent magnitude cutoff of r = 21.6. While DECaLS technically goes as deep as r = 23, after applying a photometric redshift uncertainty cutoff of 0.04, very few galaxies fell above a magnitude of 21.6 due to the fact that fainter galaxies had higher redshift errors, as can be seen in Figure \ref{fig:2dhists}. In Figure \ref{fig:maglim}, the absolute magnitudes are plotted for our real sample of FoF-selected group galaxies and the bent/unbent host galaxies as a function of redshift, with the blue line representing the absolute magnitude for a galaxy of apparent magnitude r = 21 at various redshifts.

Once a final catalog of simulation subhalos was created, we then updated the group catalog to only consider subhalos that were in our final catalog, and set another cutoff such that our FoF algorithm only considered groups in which there were at least 4 subhalos that fulfilled the criteria outlined above. This allowed us to exclude groups of 3 or fewer that may have had large spatial separations and thus would not have been picked up by any FoF algorithm.

\subsection{Testing Group Finding Algorithm Parameters}

\begin{figure}
    \centering
    \includegraphics[scale=.6]{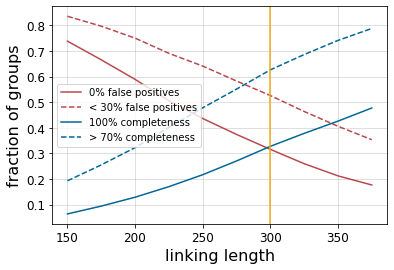}
    \includegraphics[scale=.6]{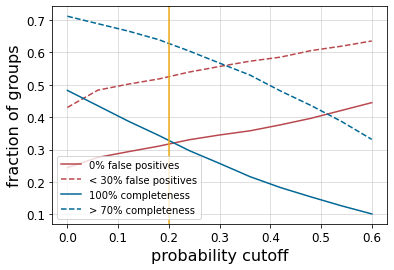}
    \includegraphics[scale=.6]{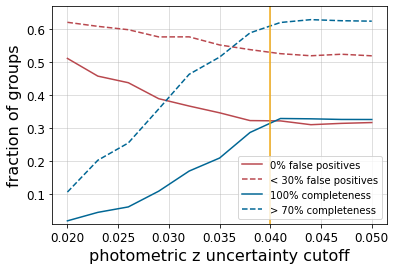}
    \caption{Results of FoF testing in Illustris when varying the linking length (top), probability cutoff (middle), and photometric redshift uncertainty cutoff (bottom). The solid gold vertical lines represent the parameter value selected for the FoF algorithm in all three cases. The blue lines show how the fraction of groups with 100$\%$ completeness (solid) and over 70$\%$ completeness change as the parameter is varied, while the red lines show how the fraction of groups with 0$\%$ false positives (solid) and less than 30$\%$ false positives (dashed) change as the parameter is varied. Parameters were selected by balancing the completeness and false positive rate as much as possible. In these plots, the parameters that were held constant were the final parameters while the third was varied.}
    \label{fig:hell}
\end{figure}

\begin{figure}
    \centering
    \includegraphics[scale=.5]{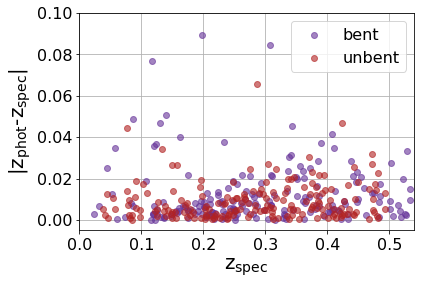}
    \caption{The difference between the photometric redshift measured by \cite{zou2019} and the spectroscopic redshift measured by SDSS as a function of the spectroscopic redshift.}
    \label{fig:zdiffs}
\end{figure}

\begin{figure}
    \centering
    \includegraphics[scale=0.6]{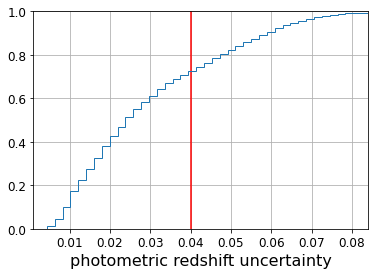}
    \caption{A cumulative histogram of photometric uncertainties from DECaLS. The red line is drawn at an uncertainty of 0.04, which is the redshift uncertainty cutoff for the FoF algorithm. This winds up including 70\% of the galaxies in DECaLS, with the rest falling above that cutoff.}
    \label{fig:cumulativeunc}
\end{figure}

The parameters that had to be tested and tuned for the group finding algorithm are: the linking length, the photometric redshift uncertainty cutoff, and the cutoff for the probability that two galaxies are at the same redshift. To tune these parameters, we carried out a series of tests in which, for each group in the Illustris group catalog, we selected a random group member and ran the group finding algorithm starting at that galaxy. We then compared the results of the group finding algorithm to the real group catalog in order to determine how well the algorithm did.

It was essential to select the parameters that resulted in groups that minimize the number of interlopers and maximize the completeness. Therefore, we specified and measured a couple of quantities for each group. The first, dubbed the false positive rate, is defined as the percentage of galaxies, out of those identified by the algorithm, that were not real group members. The second, the completeness, is defined as the percentage of galaxies, out of the real group members, that were correctly detected by the algorithm. These values tend to be correlated with one another since, if the algorithm considers more galaxies as group members, it will be able to correctly identify more group galaxies that may be on the edges of the group, but it will also likely identify more false positives, as well. Therefore, it is important to strike a reasonable balance between these two effects when calibrating parameter values.

We ran the group finding algorithm on the entire group catalog multiple times to understand how each parameter affects the accuracy of the group finding results. To begin with, we selected initial values for each parameter and measured the completeness and false positive rate for each group. Then, we began to vary one parameter at a time and looked at how the distribution of completeness and the false positive rate for all of the groups changed, selecting the one that had the best balance between the two. We would then update that parameter value, keep it static, and begin varying the next parameter. This process was repeated for each of the three parameters until we found a set of values that maximized completeness while minimizing false positives. Then, we started over again, only now using the updated parameters as initial static values and then varying one parameter at a time to see if there were any changes. Figure \ref{fig:hell} shows how the completeness and false positive rate changed as we varied each parameter, with the solid gold lines indicating the final choice of parameter.

Each of the three parameters that we varied came with its own unique set of challenges, which are summarized in the subsections below.

\subsubsection{Photometric Redshift Uncertainty}

The biggest downside to performing a FoF algorithm on data that uses photometric redshift measurements as opposed to spectroscopic ones is that these tend to have much higher uncertainties due to the fact that they rely heavily on SED-fitting of broadband filters. Because of this, we imposed an uncertainty cutoff for our photometric redshift error measurements. Before we carried out tests with Illustris, however, we began by examining how different uncertainty cutoffs affect how many galaxies we exclude from this analysis. This helped us decide on a range of values over which we performed testing with Illustris.

Comparisons between DECALS photometric redshifts and SDSS spectroscopic redshifts can be found in Figure \ref{fig:zdiffs}, where we can see the overall scatter increase at higher redshifts. We also looked at the distribution of photometric redshift uncertainties for DECaLS galaxies, which can be seen in Figure \ref{fig:cumulativeunc}, to understand how imposing a redshift cutoff would impact our final sample. Ideally, we would only be using galaxies whose photometric redshift uncertainties were low to minimize the number of false positives identified by the algorithm. However, too low of a cutoff would leave out a large percentage of galaxies in DECaLS. Therefore, we decided to run the FoF tests on Illustris while varying the uncertainty cutoff between 0.02 and 0.05.

The results of the final test are summed up by Figure \ref{fig:hell}, which shows how these quantities vary over a range of uncertainty cutoffs. Figure \ref{fig:cumulativeunc} shows the cumulative distribution of photometric redshift uncertainties for galaxies observed in DECaLS. We selected a redshift cutoff of 0.04 in the end due to the fact that it provides a good balance between accuracy of the FoF tests without excluding too many galaxies from our analysis. This ends up including roughly 70$\%$ of the galaxies in DECaLS.

\subsubsection{Linking Length} \label{sec:caliblinlen}

Next, we needed to calibrate the linking length needed for our analysis. It is important to note that, for our analysis, we want to consider as many galaxies near our AGN as possible so we can get an accurate idea of the environments they are in. Additionally, the fact that we are using photometric redshifts as opposed to spectroscopic redshifts means that scaling the linking length by the number density of galaxies in our sample, as many other studies using spectroscopic redshifts have done, will be difficult. Therefore, we use a static value for the linking length that has been tuned using Illustris TNG300.

In order to find the best value for the linking length, we measure the distance to the nearest galaxy for each simulation subhalo in a group. This distribution can be found in Figure \ref{fig:nearneigh}, where we have also plotted the mean and median for the distribution, as well as what those values would be in 2D space. This leaves us with 4 potential values for the linking length. We ran the FoF algorithm on the constructed Illustris catalog projected onto the sky using linking lengths that spanned from 190 to 320 kpc, which roughly coincided to the lowest and highest values shown in Figure \ref{fig:nearneigh}. We chose a value of 300 kpc to strike a balance between including a large fraction of group members without falsely identifying too many galaxies not associated with the group as group members.

\subsubsection{Probability Metric} \label{sec:pickprob}
Because our particular FoF algorithm only uses a probability-based metric along the line of sight, we must be sure that this metric results in both complete and accurate groups. In order to do this, we ran the FoF algorithm on the simulation groups using minimum probability metrics that ranged from 0 (i.e. no redshift consideration) to 0.8. We found that a probability metric of 0.2 resulted in the best balance between low false positive rate and a high completeness, making this the value we decided to use for the probability metric.

\begin{figure*}
    \centering
    \includegraphics[scale=.6]{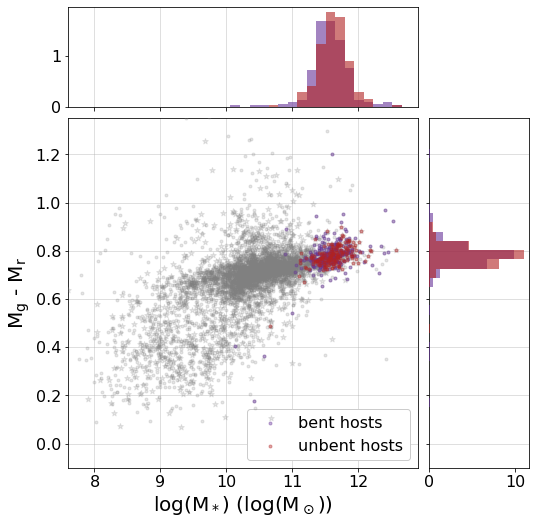}
    \caption{A color-mass plot using k-corrected absolute magnitudes from DECaLS of unbent and bent doubles, along with galaxies identified as group members.}
    \label{fig:colormass}
\end{figure*}

\begin{deluxetable}{ccccc}
\tablenum{1}
\tablecaption{AGN Environment Summary}
\tablewidth{0pt}
\tablehead{
\colhead{AGN}& \colhead{$\#$ of} & \multicolumn{3}{c}{\% of AGN...}\\[-0.2cm]
\colhead{Morphology/}& \colhead{AGN} & \colhead{in group,} & \colhead{in group,} & \colhead{in pair}\\[-0.2cm]
\colhead{Environment}& \colhead{} & \colhead{not brightest} & \colhead{brightest} & \colhead{or isolated}}
\startdata
Bent, all & 175 & 21.1\% & 62.3\% & 16.6\% \\
Bent, groups& 150 & 25.3\% & 74.7\% & - \\
Unbent, all & 187 & 5.9\% & 66.3\% & 27.8\% \\
Unbent, groups& 135 & 8.1\% & 91.9\% & - \\
\enddata
\tablecomments{The second and fourth rows only consider the AGN in groups of three or more and therefore do not have values in the third column, which counts the number of AGN with 2 or fewer neighbors.}
\end{deluxetable}

\section{Results}

A color-mass diagram is plotted in Figure \ref{fig:colormass} to showcase general characteristics of the radio galaxies and galaxies that reside in the same halos as them. Unsurprisingly, for the most part, both the bent and unbent AGN hosts reside in massive, red galaxies. This fits with the unified model of AGN in which accretion flows that launch radio jets exist in environments similar to the centers of massive elliptical galaxies \citep{heckman2014}.

We also see an overdensity of group members that reside in the red cloud, indicating that these galaxies are either quenched or in dust-obscured systems. This is not unexpected, as galaxies in dense environments (such as the groups or clusters that many of the radio galaxies in our sample are probing) are more likely to be red and quenched early-type galaxies.

\begin{figure}
    \centering
    \includegraphics[scale=.55]{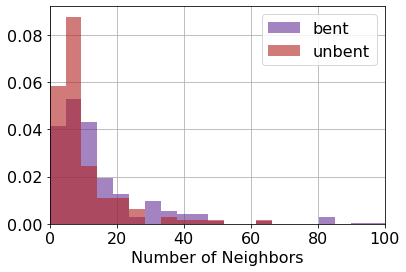}
    \caption{Normalized distributions of the number of galaxies occupying the same groups as bent and unbent radio galaxies. This distribution only includes groups with at least 3 members and excludes all groups with more than 100 members. This results in us excluding 7 groups hosting bent doubles and 2 groups hosting unbent doubles.}
    \label{fig:nneigh}
\end{figure}

\begin{figure}
    \centering
    \includegraphics[scale=0.6]{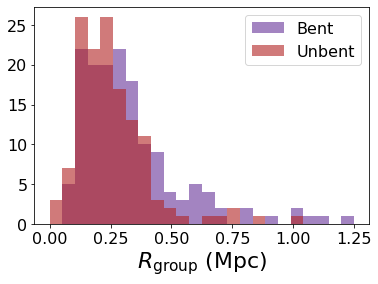}
    \caption{Comparison of the radii of groups hosting bent and unbent radio AGN.}
    \label{fig:rad}
\end{figure}

\subsection{The Frequency of AGN in Different Environments}

We look into the properties of bent and unbent environments by comparing how often the galaxies hosting AGN are alone or in pairs, if they are the brightest galaxy in their group with 3 or more galaxies, or if they are in a group with 3 or more galaxies and are not the brightest galaxy. This information is summarized in Table 1. We choose these three categories to first identify how many galaxies are in underdense environments, such as singles or pairs. We then separate groups of 3 or more galaxies into groups where AGN are the brightest and groups where the AGN is not the brightest. If the AGN host is the Brightest Group Galaxy (BGG), it is very likely that it has undergone an evolution defined largely by hierarchical structure formation of merging galaxies \citep{delucia2007}, where it has been able to grow more massive than a typical large elliptical galaxy by accreting satellite galaxies that fall into the gravitational potential of the group \citep{gozaliasl2019,pasini2021}.

One of the most significant differences between the bent and unbent populations shown in Table 1 is the percentage that are alone or in pairs versus in groups of three or more galaxies. We looked at pairs and singles because galaxies in these environments are likely not embedded in a dense IGM like those in groups of three or more \citep{garon2019}. In our study, 16.6\% of bent doubles are in pairs, while 27.8\% of unbent doubles are in pairs. Therefore, while it is unlikely to find a bent AGN alone or in a pair, they do exist, just not to the extent that unbent AGN do.

While the percentage of groups where the radio galaxies are the brightest is similar between the bent and unbent samples, a larger difference arises when looking at how many AGN hosts are alone/in pairs, or are not the BGG. These numbers become more stark when only considering galaxies in groups of 3 or more members. In 25.3\% of the groups with bent AGN, the AGN host galaxy was not the brightest group galaxy, while this number was reduced to 8.1\% when considering unbent AGN. This shows that AGN with unbent morphologies are less likely than bent AGN to exist in groups of three or more galaxies without being the BGG.

We performed a Pearson's chi-squared test to the data in Table 1 in order to see if the differences noted above are real, or if they could have been observed by chance. We found that we can reject the null hypothesis that this was a chance observation of the same population with a confidence level of over 99.99$\%$ and a p-value of $10^{-5}$.

To summarize these comparisons, unbent AGN are more likely to exist alone or in a pair than bent AGN. When in groups of three or more galaxies, unbent AGN are very likely to be the brightest cluster galaxy. For bent AGN, while they are still fairly likely to be the BGG, there is still a large population of bent AGN that exist in groups without being the BGG.

\begin{figure}
    \centering
    \includegraphics[scale=.6]{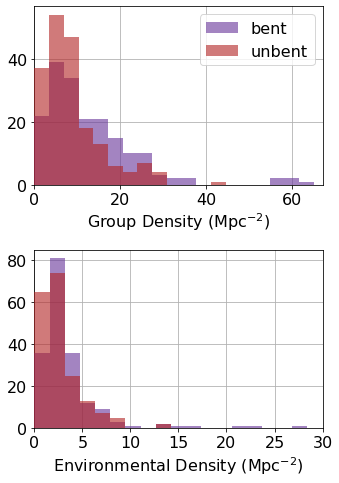}
    \caption{Density measures for groups hosting bent (purple) and unbent (red) radio galaxies. (Top) The group density, measured within 0.5 Mpc of the group center. (Bottom) The environmental density, measured from 0.5 Mpc to 2 Mpc of the group center.}
    \label{fig:density}
\end{figure}

\begin{figure}
    \centering
    \includegraphics[scale=0.5]{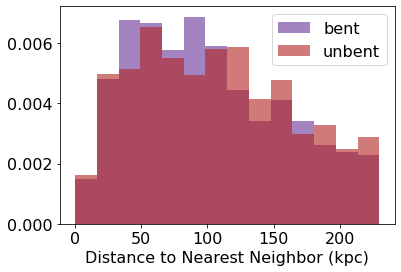}
    \caption{A normalized histogram of distance to nearest neighbors using the results of the FoF algorithm on the DECaLS data.}
    \label{fig:dnn}
\end{figure}

\subsection{Group Membership}
When we refer to the size of the group, we are referring to the number of group members identified by the FoF algorithm. The distribution of group membership for groups of three or more galaxies can be seen for groups occupied by both bent and unbent radio galaxies in Figure \ref{fig:nneigh}. The overall shapes of these distributions is somewhat similar, but it is clear that the unbent sample is skewed towards lower membership, while the bent sample reaches higher membership. Using the k-sample Anderson Darling test, we find that the null hypothesis can be rejected at a confidence level of 99.99\%, indicating that these are two distinct populations.

\subsection{Group Radius}

We estimate the projected group radius using the equation below.

{\bf $$R^2_{\text{group}} = \frac{1}{n}\sum_{i = 1}^{n} r_i^2$$}

This is the mean square of $r_i$, the comoving distance between a galaxy $i$ to the group center calculated using astropy's cosmology package \citep{astropy:2013,astropy:2018}. This equation is similar to that used by \cite{tempel2014} to estimate the group extent. Distributions of radii for groups hosting bent and unbent radio AGN can be found in Figure \ref{fig:rad}, which shows that bent AGN are more likely to exist in groups with larger radii. We performed an Anderson-Darling test to determine the likelihood that these two distributions could be derived from the same parent population and found that we could reject the null hypothesis to a confidence level of 3-$\sigma$, obtaining a p-value of less than 0.002. It is worth noting that the actual difference in radius of these two samples is fairly small, with the average group hosting bent AGN having a radius of 332 kpc and the average group hosting an unbent AGN having a radius of 259 kpc. It is also worth noting that the majority of all groups in our sample have radii below 500 kpc, providing quantitative justification for using 500 kpc as the radius within which we use to calculate the group density.

\subsection{Group Density}

We use multiple proxies as tracers of the density in galaxy groups: the number density of galaxies projected within 0.5 Mpc, between 0.5 and 2 Mpc of the group center, and the distance to the nearest group member. These distances are measured in the plane of the sky and we only consider galaxies with photometric redshift measurements that meet the probability criterion outlined in Section 3.2. By using these static values as opposed to the density to the nth nearest neighbor, we are better able to probe a wide range of environments, from pairs of galaxies to groups that are on the verge of being clusters. Additionally, these measurements are less affected by potential false positives that could be identified by our algorithm.

The maximum projected distance a group member can be found from the bent double is 2 Mpc due to the constraints placed on the algorithm (outlined in Section 3.1.1. and 4.3). However, there are very few groups that actually reach this cutoff, and most groups do not extend too far beyond 0.5 Mpc \citep{tempel2014,riggs2021}. Therefore, when measuring the galaxy density of the core of the group, we consider all galaxies within 0.5 Mpc of the group center. This is partially motivated by the richness metric used by various studies in which the number of galaxies below an absolute magnitude threshold within 0.5 Mpc of a galaxy cluster or group center are counted \citep{allingtonsmith1993,zirbel1997,wingblanton2011}. A more recent study of clustering in galaxy groups found that most galaxy groups reach peak clustering at a radius of 0.5 Mpc, beyond which, the clustering falls off \citep{riggs2021}. For the galaxy density of the surrounding environment, we consider all galaxies between 0.5 and 2 Mpc, as galaxies at these distances from the group center are less likely to be gravitationally coupled to the group.

Figure \ref{fig:density} shows the results of our group finding algorithm for groups hosting both bent and unbent galaxies. When considering the density of the inner 0.5 Mpc of the group, there is a clear distinction between bent and unbent doubles: bent doubles are more likely to reside in halos that are denser than their unbent counterparts. We performed a series of statistical tests in order to ensure that these two histograms are not being drawn from the same sample. Using the Anderson-Darling k sample test, we found that the null hypothesis that these two samples are drawn from the same population can be rejected with a statistical significance of 0.1$\%$. We also used the Mann-Whitney-Wilcoxon Test, which tests the location of separate distributions and compares them to see how similar the distributions themselves are. This test rejected the null hypothesis that these two distributions are the same with a reported p-value of less than 0.01. These separate statistical tests give us a reasonable degree of confidence when stating that bent doubles are indeed found in more dense groups than unbent doubles.

When considering the density of the larger environment, the difference between bent and unbent doubles is less stark, but still present. We can reject the null hypothesis using the Anderson-Darling test at roughly 1.1$\%$ significance and using the Mann-Whitney-Wilcoxon Test with a p value of less than 0.01. This shows that even the larger environments, out to 2 Mpc, are more dense for bent AGN, if only slightly.

We also examine the distances to the nearest neighbor for galaxies in groups. Measuring the distance to the nearest neighbor for every galaxy in the group is another way to learn about the group environment and how densely packed galaxies in this environment are to one another. This is a proxy for the number density of galaxies in these groups, but it considers every galaxy in the group and is not bound by the limitations of selecting a static aperture size within which the density is measured. This means that calculating the nearest neighbor for every galaxy in the group will include group members that exist beyond 0.5 Mpc from the center of the group, making it a proxy for the density of the group as a whole instead of just the core of the group where we expect it to be at its most dense. However, it is important to note that this metric is particularly sensitive to false positives and incompleteness, as a single false positive or missed group galaxy can throw off the nearest neighbor distance for many group members. Therefore, while this metric can be another proxy for the group density, it is important to note that these caveats may lead to it not being the most accurate proxy in this study.

The results of measuring the distances to the nearest neighbor for the entire sample can be found in Figure \ref{fig:dnn}. There is a great deal of overlap between the two distributions and, while both the mean and median distance to the nearest neighbor for the bent sample are lower than the unbent sample, it is typically by 10-20 kpc. However, according the the Mann-Whitney and Anderson-Darling tests, the null hypothesis can be rejected at a statistically significant level with a p-value less than 0.01 for the Mann-Whitney test and a significance of roughly 0.7$\%$ for Anderson-Darling. It is important to note that the significance level of this measurement is not as high as those of the group densities, mainly because this particular metric is very sensitive to false positives and incompleteness of the group sample. Therefore, it is important to take the results of this metric with a grain of salt. That being said, these results are consistent with the results from the densities discussed earlier: bent doubles tend to exist in groups that have higher number densities than do unbent doubles.

\begin{figure}
    \centering
    \includegraphics[scale=.6]{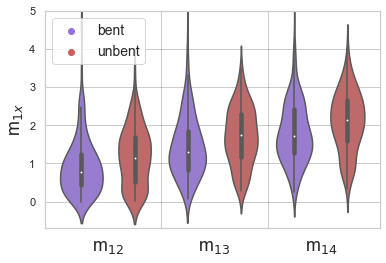}
    \caption{Violin plots showing the Gaussian KDE for three different versions of the r-band magnitude difference. (Left) Magnitude difference between the brightest and second brightest group galaxies, (middle) difference between brightest and third brightest group galaxies, (right) difference between the brightest and fourth brightest group galaxies. For all three metrics, the unbent sample sits at a higher magnitude difference.}
    \label{fig:rdiff}
\end{figure}
\subsection{Magnitude Gap}

Because we are working with less precise photometric redshifts, we are unable to precisely measure the group dynamics. However, we can use the r-band magnitude gap to understand the dynamical history and age of the groups (discussed further in Section 5.1). The r-band magnitude difference between the first and second brightest group galaxies can reveal something about the merger history of the group, and thus its dynamical age. Because galaxy groups are formed via hierarchical structure formation, the BGG will tend to gain mass via mergers with satellites \citep{donghia2005,delucia2007,oogi2016}. This evolution leads BGGs to be statistically more massive, and thus brighter than the next brightest group galaxy, and has been shown to correlate with the r-band magnitude gap \citep{yang2008,shen2014,solanes2016,dalal2021}. In this study, we use the r-band magnitude gap as a proxy for the dynamical ages of the galaxy groups in our sample.

In Figure \ref{fig:rdiff}, we plot the Gaussian Kernel Density Estimates (KDEs) of the r-band magnitude differences using the difference between the first and second (m$_{12}$), the first and third (m$_{13}$), and the first and fourth (m$_{14}$) brightest group galaxies. Our justification for using multiple metrics is both because this is what has been used in the literature \citep{solanes2016} and to help offset any issues introduced by false positives. For all three metrics, the median magnitude gap is higher for the unbent sample by roughly 0.4 mag, and after applying the Anderson-Darling test, we can reject the null hypothesis that these distributions are the same to a statistical significance level of roughly 0.2$\%$. The Mann-Whitney test corroborates this with p-values of roughly 10$^{-3}$. These results indicate that bent AGN tend to exist in groups with lower magnitude gaps than their unbent counterparts.

\subsection{How do group finding parameters affect the overall results?}

One of the downsides to using an FoF algorithm is the fact that the linking length is generally a difficult parameter to tune, and can result in very different looking environments depending on the choice. Additionally, our group finding algorithm requires that we set a cutoff for the photometric redshift and a cutoff for the relative probability that two galaxies are at the same redshift, for a total of three selectable parameters. While the parameters were chosen in a systematic manner to optimize the balance between completeness and false positive rate, it is prudent to examine the robustness of the main results across a range of parameter values.

We performed the group finding algorithm on the real data two additional times: once with all of the parameters increased and once with them decreased. We varied the linking length by 100 kpc, the z uncertainty cutoff by 0.01, and the probability cutoff by 0.15. Tuning these parameters to their extreme values should result in the largest differences between our test groups and the groups we get using the original parameter values. With these test groups, we carried out the same statistical analyses on distributions of group membership, group and environmental density, and magnitude differences.

For both tests, we found that, even with the parameters at different values, the overall results stay the same. The only difference is that the results are not quite as statistically significant. The largest difference is that the magnitude gap in the low parameter case goes from a significance level of 0.2$\%$ to 3.4$\%$. So, even in the most extreme cases, the statistical significance only changes by a few percent, indicating that our results are robust.


\section{Discussion}

In this study, we have compared the environments of bent and unbent double lobed radio AGN in order to learn more about what could be causing the bending of bent doubles and why we do not see the same effect on unbent AGN. An AGN's radio jets could be bent by the ram pressure of the surrounding IGM \citep{begelman1979,morsony2013}. How much the jets are bent depends on the density of the gas through with the AGN host galaxy is moving and the velocity of the AGN host galaxy relative to the IGM. It is also possible that a recent group or cluster merger could cause the IGM to become more turbulent and provide the ram pressure necessary to bend the jets out to large distances from the cluster center \citep{sakelliou2000,blanton2001,devos2021}. Therefore, we would expect to see these kinds of galaxies existing in denser environments and moving much more rapidly compared to their unbent counterparts. Our results indicate that this could, indeed, be the case for this sample of bent doubles.

From this work, we can say to a statistically significant degree that AGN with bent jets are more likely than unbent AGN to exist in groups of 3+ members, tend to exist in larger, denser, and richer environments, and tend to exist in environments with lower r-band magnitude gaps than unbent galaxies, indicating that they exist in dynamically younger environments that may have undergone a recent merger. These results correspond well with the results of \cite{garon2019} and \cite{devos2021}, which showed that, while bent AGN tend to exist in fairly dense environments, they are not always located directly towards the center of the nearest group or cluster and instead can exist in different kinds of environments.

Measuring the density of the IGM in groups directly is very difficult, due to the fact that it likely exists in a hard-to-observe state known as the warm-hot intergalactic medium \citep[WHIM;][]{cenostriker2006}, which typically exists at 10$^5$ - 10$^7$ K. While there have been previous X-ray \citep[][]{sanderson2013} and Sunyaev-Zel'dovich effect \citep[SZ effect;][]{lim2017} measurements of hot gas in galaxy groups, both of these fall below the expected cosmological baryon fraction in groups. This is potentially due to the fact that current X-ray and SZ effect observations only probe the hottest gas in these environments and are not sensitive to cooler gas that is difficult to observe. Because of this, we use the number density of galaxies themselves as a proxy for the density of the IGM gas, which is ultimately what is responsible for bending AGN jets under the assumption of ram pressure.

There could potentially be other reasons for AGN jets being bent, such as interactions with nearby galaxies or the interactions between the AGN jet and the ISM within the host galaxy causing bent jet morphologies, such as sharp bending near the host galaxy. Precession of the SMBH could also cause twisting bent jet morphologies that could result in an S-shaped jet \citep{falcetagoncalves2010,roberts2018}. However, we attempt to take these into account when visually selecting bent AGN, only looking for those with smooth bending in one direction.

Based on our visual selection, most of the galaxies in our sample would be considered WATs. These types of bent AGN are typically found in the BGG/BCG \citep{sakelliou2000,blanton2001}. However, there are other recent studies that have found that they can exist outside of the nearest group or cluster, potentially indicating that they may be embedded in a filamentary structure or that the impact of the cluster/group medium is widespread \citep{garon2019,devos2021}. In this study, we found that 62.3$\%$ of bent doubles are the BCG/BGG. This is consistent with the value of 55$\%$ from \cite{goldenmarx2021}, who studied mostly WATs in cluster environments. The fact that this fraction is as high as it is may be in part due to the visual sample largely being made up of WATs as opposed to NATs, which generally do not exist in the BCG/BGG. Additionally, BCGs/BGGs already have a high likelihood of being bright radio galaxies in the first place, making them more easily detectable as bent doubles. However, it is still worth noting the large fraction of bent doubles that are either isolated, in pairs, or in groups without being the BCG/BGG.

One effect we are unable to take into account is the projection of the jets along the line of sight. The only way we are able to observe a jet as bent is if the relative motion between the galaxy and the IGM has a substantial component perpendicular to the line of sight. If it is directed towards or away from us, however, we would be unable to directly see the bending in the radio observations. Additionally, the viewing angle of the jet can affect how it appears to us. If the viewing angle is close to 0$^\circ$ and the jets are pointed directly at us, then it would be difficult to discern any kind of bending due to relativistic beaming or appear to be much more bent than they are. Alternatively, if the viewing angle is less than 90$^\circ$ while still being able to resolve two separate lobes, the jets will appear to be more bent, potentially making a WAT look like a NAT. A deeper discussion of how projection affects the morphology we assign a radio galaxy is beyond the scope of this paper.

\subsection{Dynamical histories of groups with bent and unbent AGN}

Hierarchical structure formation drives much of the evolution of BCGs and BGGs, causing them to grow in mass and luminosity, especially compared to the next brightest galaxies in the group \citep{donghia2005,delucia2007,oogi2016}. This leads to the r-band magnitude gap between first and second brightest galaxy growing over time as the BCG continues to accrete mass and luminosity \citep{yang2008,shen2014,solanes2016,dalal2021}. Therefore, the r-band magnitude gap can be used as a tracer for the dynamical age and the current dynamics of a cluster or group \citep{shen2014,lopes2018}, with larger magnitude gaps indicating groups that are more settled and whose BGGs have had more time to accrete mass. On the other hand, smaller magnitude gaps could indicate that a group has recently formed or merged with another group, which could lead to the group being less dynamically relaxed and could potentially result in turbulence in the IGM or galaxies with higher velocities, making conditions ideal for AGN jets to bend. This could also indicate a recent merger may have caused the BGG to be offset from the group center \citep{sakelliou2000}. The fact that 62.3$\%$ of bent doubles are the BGG could indicate that they exist in recently merged systems and are moving through a dense, turbulent medium towards the new group center. It is worth noting, however, that the findings of \cite{wingblanton2013} indicate that there is not a significant difference in the substructure of clusters hosting bent and unbent doubles, supporting the idea that cluster-cluster mergers are not solely responsible for bending AGN jets.

We find that, using a variety of metrics, groups hosting AGN with bent jets have lower r-band magnitude gaps, implying that they exist in groups that have formed more recently and whose BGGs have not had the time to grow appreciably due to hierarchical structure formation, or that they are part of groups that have recently merged, causing the two brightest galaxies to be similar in luminosity.

\subsection{Bent doubles in singles and pairs}

A dense environment is likely to play an important role in bending the jets for the most part, but there is still a number of AGN with bent jets in our sample that seemingly exist outside of groups or clusters. This is consistent with with the results of \cite{garon2019}, who found that bent AGN are not necessarily always located near known clusters, but still tend to exist in overdensities of 28$\rho_{\text{crit}}$ (highly bent) and 18$\rho_{\text{crit}}$ (less bent) from 50 - 350 kpc. \cite{wingblanton2011} also found that, while most galaxies hosting bent AGN existed in overdense regions, there were still a few that existed in environments with very few nearby neighbors. \cite{edwards2010} found a bent double located in a cosmic web filament and was able to use it to measure the density of gas in the filament and \cite{devos2021} found that NAT radio sources can exist out to 10$R_{500}$ of a nearby cluster. It could be that many of these bent doubles in singles and pairs are on the outskirts of a larger group or cluster and are embedded in a filament similar to that found in \cite{edwards2010} or those observed by \cite{devos2021}. Deeper, more complete spectroscopic observations of isolated bent doubles and any potential neighboring galaxies could provide us with a more complete picture of their environments, helping to better reveal the causes bending in very sparse environments.

\subsection{Unbent doubles that are not the BGG/BCG}
We find that 5.9$\%$ of unbent AGN in groups are not the BCG/BGG. There is evidence that a radio galaxy's position in a group/cluster depends on factors of the radio source itself, such as the radio luminosity \citep{croston2019}, with fainter radio galaxies existing further from the group/cluster center. Studies have also found AGN with unbent jets outside of the brightest cluster galaxy, even in higher redshift clusters \citep{moravec2020}. The fact that these galaxies are not bent might appear surprising, as we would expect a group/cluster member that is not the brightest galaxy to likely be moving through the group or cluster and thus be susceptible to ram pressure from the surrounding medium that would lead to bending. It could be possible that these particular galaxies are in environments with lower IGM densities, or that the jets are powerful enough to overcome the influence of ram pressure from the surrounding medium. Additionally, bending should decrease at higher distances from the cluster center, so the fact that there are some galaxies that are unbent and not located at the center of the cluster is not surprising \citep{garon2019}. There is also a chance that these galaxies could actually host bent jets that are moving directly towards or away from us such that we are unable to see the bending.


\section{Conclusions}
Using a group finding algorithm that combines FoF methods and probabilistic methods, we performed a study of the environments surrounding AGN with radio jets that are either bent or unbent. For an AGN's jets to bend, its host galaxy is likely moving through a dense reservoir of gas whose ram pressure is causing the jets to bend back. However, there are also AGN with unbent jets that exist in dense environments. This study compared the environments around samples containing these two kinds of galaxies in order to better understand why some galaxies have bent AGN jets and others do not.

The main takeaways of this study are listed below:

\begin{itemize}
    \item The results of our group finding algorithm show that 16.6\% of the bent double host galaxies in our sample are in singles or in pairs, while this number rises to 27.8\% for unbent doubles. This indicates that unbent galaxies are more likely to exist in less dense environments where their jets will not necessarily interact with the surrounding medium enough to be bent, but that there is a subset of bent doubles that also exist in this environment. What could be responsible for bending their jets is yet to be determined, but they could be in small overdensities of the IGM or in cosmic web filaments.
    
    \item The results of the group finding algorithm also show that 25.3\% of bent doubles in groups of at least 3 galaxies are not the brightest galaxy, while this number is 8.1\% for unbent doubles. Bent doubles are less likely to be the BGG in the centers of groups, meaning that more of them may be satellites moving quickly through the IGrM.
    
    \item In groups of 3 or more members, bent doubles are more likely to exist in halos with a higher number of galaxies.
    
    \item Groups hosting bent AGN tend to have higher radii than those hosting unbent AGN.
    
    \item The inner 0.5 Mpc of groups with bent doubles tend to be more dense than those with unbent doubles. This could mean that there is also more gas in the IGrM that is causing these jets to bend.
    
    \item Groups hosting bent doubles tend to have lower r-band magnitude gaps than groups with unbent doubles. This could be an indication that groups with bent doubles are less dynamically relaxed or are younger than groups with unbent doubles.
    
    \item Of all unbent doubles, 5.9\% exist in groups without being the brightest group galaxy. This could be due to projection
\end{itemize}

Further improving our understanding of the differences between the environments inhabited by bent and unbent doubles would require additional information. First of all, more accurate and complete spectroscopy of galaxy group members or neighboring galaxies would allow for a more complete census of the environments surrounding the radio galaxies, as well as the dynamical state of their environments. Furthermore, deep X-ray observations of these groups would allow us to probe for a hot gaseous intergalactic medium that may be responsible for bending some of the AGN. eROSITA will likely provide important constraints for the galaxies in our sample and beyond.

Additionally, upcoming deep and high resolution radio surveys such as the VLA Sky Survey \citep[VLASS;][]{lacy2020}, the LOFAR Two-metre Sky Survey \citep[LoTSS;][]{shimwell2017}, the Meerkat Galaxy Cluster Legacy Survey \citep[MGCLS;][]{knowles2022}, and eventually the Square Kilometre Array (SKA) will provide imaging of even more bent and unbent radio AGN jets, giving us a much larger sample of galaxies to work with. Machine learning techniques could be developed to aid in the morphology classification of these sources, allowing the number of bent AGN detected to increase by a large margin.

\begin{acknowledgments}

Support for this research was provided by the University of Wisconsin-Madison, Office of the Vice Chancellor for Research and Graduate Education with funding from the Wisconsin Alumni Research Foundation. We acknowledge support from NSF AST-1616101. We would like to thank the anonymous referee and the statistics editor for their helpful comments and suggestions.

\software{Astropy \citep{astropy:2013, astropy:2018}}

\end{acknowledgments}

\bibliography{mybib}{}

\begin{thebibliography}{}
\expandafter\ifx\csname natexlab\endcsname\relax\def\natexlab#1{#1}\fi
\providecommand{\url}[1]{\href{#1}{#1}}
\providecommand{\dodoi}[1]{doi:~\href{http://doi.org/#1}{\nolinkurl{#1}}}
\providecommand{\doeprint}[1]{\href{http://ascl.net/#1}{\nolinkurl{http://ascl.net/#1}}}
\providecommand{\doarXiv}[1]{\href{https://arxiv.org/abs/#1}{\nolinkurl{https://arxiv.org/abs/#1}}}

\bibitem[{{Abazajian} {et~al.}(2009){Abazajian}, {Adelman-McCarthy},
  {Ag{\"u}eros}, {Allam}, {Allende Prieto}, {An}, {Anderson}, {Anderson},
  {Annis}, {Bahcall}, {Bailer-Jones}, {Barentine}, {Bassett}, {Becker},
  {Beers}, {Bell}, {Belokurov}, {Berlind}, {Berman}, {Bernardi}, {Bickerton},
  {Bizyaev}, {Blakeslee}, {Blanton}, {Bochanski}, {Boroski}, {Brewington},
  {Brinchmann}, {Brinkmann}, {Brunner}, {Budav{\'a}ri}, {Carey}, {Carliles},
  {Carr}, {Castander}, {Cinabro}, {Connolly}, {Csabai}, {Cunha}, {Czarapata},
  {Davenport}, {de Haas}, {Dilday}, {Doi}, {Eisenstein}, {Evans}, {Evans},
  {Fan}, {Friedman}, {Frieman}, {Fukugita}, {G{\"a}nsicke}, {Gates},
  {Gillespie}, {Gilmore}, {Gonzalez}, {Gonzalez}, {Grebel}, {Gunn},
  {Gy{\"o}ry}, {Hall}, {Harding}, {Harris}, {Harvanek}, {Hawley}, {Hayes},
  {Heckman}, {Hendry}, {Hennessy}, {Hindsley}, {Hoblitt}, {Hogan}, {Hogg},
  {Holtzman}, {Hyde}, {Ichikawa}, {Ichikawa}, {Im}, {Ivezi{\'c}}, {Jester},
  {Jiang}, {Johnson}, {Jorgensen}, {Juri{\'c}}, {Kent}, {Kessler}, {Kleinman},
  {Knapp}, {Konishi}, {Kron}, {Krzesinski}, {Kuropatkin}, {Lampeitl},
  {Lebedeva}, {Lee}, {Lee}, {French Leger}, {L{\'e}pine}, {Li}, {Lima}, {Lin},
  {Long}, {Loomis}, {Loveday}, {Lupton}, {Magnier}, {Malanushenko},
  {Malanushenko}, {Mandelbaum}, {Margon}, {Marriner}, {Mart{\'\i}nez-Delgado},
  {Matsubara}, {McGehee}, {McKay}, {Meiksin}, {Morrison}, {Mullally}, {Munn},
  {Murphy}, {Nash}, {Nebot}, {Neilsen}, {Newberg}, {Newman}, {Nichol},
  {Nicinski}, {Nieto-Santisteban}, {Nitta}, {Okamura}, {Oravetz}, {Ostriker},
  {Owen}, {Padmanabhan}, {Pan}, {Park}, {Pauls}, {Peoples}, {Percival}, {Pier},
  {Pope}, {Pourbaix}, {Price}, {Purger}, {Quinn}, {Raddick}, {Re Fiorentin},
  {Richards}, {Richmond}, {Riess}, {Rix}, {Rockosi}, {Sako}, {Schlegel},
  {Schneider}, {Scholz}, {Schreiber}, {Schwope}, {Seljak}, {Sesar}, {Sheldon},
  {Shimasaku}, {Sibley}, {Simmons}, {Sivarani}, {Allyn Smith}, {Smith},
  {Smol{\v{c}}i{\'c}}, {Snedden}, {Stebbins}, {Steinmetz}, {Stoughton},
  {Strauss}, {SubbaRao}, {Suto}, {Szalay}, {Szapudi}, {Szkody}, {Tanaka},
  {Tegmark}, {Teodoro}, {Thakar}, {Tremonti}, {Tucker}, {Uomoto}, {Vanden
  Berk}, {Vandenberg}, {Vidrih}, {Vogeley}, {Voges}, {Vogt}, {Wadadekar},
  {Watters}, {Weinberg}, {West}, {White}, {Wilhite}, {Wonders}, {Yanny},
  {Yocum}, {York}, {Zehavi}, {Zibetti}, \& {Zucker}}]{abazajian2009}
{Abazajian}, K.~N., {Adelman-McCarthy}, J.~K., {Ag{\"u}eros}, M.~A., {et~al.}
  2009, \apjs, 182, 543, \dodoi{10.1088/0067-0049/182/2/543}

\bibitem[{{Allington-Smith} {et~al.}(1993){Allington-Smith}, {Ellis}, {Zirbel},
  \& {Oemler}}]{allingtonsmith1993}
{Allington-Smith}, J.~R., {Ellis}, R., {Zirbel}, E.~L., \& {Oemler}, Augustus,
  J. 1993, \apj, 404, 521, \dodoi{10.1086/172305}

\bibitem[{{Astropy Collaboration} {et~al.}(2013){Astropy Collaboration},
  {Robitaille}, {Tollerud}, {Greenfield}, {Droettboom}, {Bray}, {Aldcroft},
  {Davis}, {Ginsburg}, {Price-Whelan}, {Kerzendorf}, {Conley}, {Crighton},
  {Barbary}, {Muna}, {Ferguson}, {Grollier}, {Parikh}, {Nair}, {Unther},
  {Deil}, {Woillez}, {Conseil}, {Kramer}, {Turner}, {Singer}, {Fox}, {Weaver},
  {Zabalza}, {Edwards}, {Azalee Bostroem}, {Burke}, {Casey}, {Crawford},
  {Dencheva}, {Ely}, {Jenness}, {Labrie}, {Lim}, {Pierfederici}, {Pontzen},
  {Ptak}, {Refsdal}, {Servillat}, \& {Streicher}}]{astropy:2013}
{Astropy Collaboration}, {Robitaille}, T.~P., {Tollerud}, E.~J., {et~al.} 2013,
  \aap, 558, A33, \dodoi{10.1051/0004-6361/201322068}

\bibitem[{{Astropy Collaboration} {et~al.}(2018){Astropy Collaboration},
  {Price-Whelan}, {Sip{\H{o}}cz}, {G{\"u}nther}, {Lim}, {Crawford}, {Conseil},
  {Shupe}, {Craig}, {Dencheva}, {Ginsburg}, {Vand erPlas}, {Bradley},
  {P{\'e}rez-Su{\'a}rez}, {de Val-Borro}, {Aldcroft}, {Cruz}, {Robitaille},
  {Tollerud}, {Ardelean}, {Babej}, {Bach}, {Bachetti}, {Bakanov}, {Bamford},
  {Barentsen}, {Barmby}, {Baumbach}, {Berry}, {Biscani}, {Boquien}, {Bostroem},
  {Bouma}, {Brammer}, {Bray}, {Breytenbach}, {Buddelmeijer}, {Burke},
  {Calderone}, {Cano Rodr{\'\i}guez}, {Cara}, {Cardoso}, {Cheedella}, {Copin},
  {Corrales}, {Crichton}, {D'Avella}, {Deil}, {Depagne}, {Dietrich}, {Donath},
  {Droettboom}, {Earl}, {Erben}, {Fabbro}, {Ferreira}, {Finethy}, {Fox},
  {Garrison}, {Gibbons}, {Goldstein}, {Gommers}, {Greco}, {Greenfield},
  {Groener}, {Grollier}, {Hagen}, {Hirst}, {Homeier}, {Horton}, {Hosseinzadeh},
  {Hu}, {Hunkeler}, {Ivezi{\'c}}, {Jain}, {Jenness}, {Kanarek}, {Kendrew},
  {Kern}, {Kerzendorf}, {Khvalko}, {King}, {Kirkby}, {Kulkarni}, {Kumar},
  {Lee}, {Lenz}, {Littlefair}, {Ma}, {Macleod}, {Mastropietro}, {McCully},
  {Montagnac}, {Morris}, {Mueller}, {Mumford}, {Muna}, {Murphy}, {Nelson},
  {Nguyen}, {Ninan}, {N{\"o}the}, {Ogaz}, {Oh}, {Parejko}, {Parley}, {Pascual},
  {Patil}, {Patil}, {Plunkett}, {Prochaska}, {Rastogi}, {Reddy Janga},
  {Sabater}, {Sakurikar}, {Seifert}, {Sherbert}, {Sherwood-Taylor}, {Shih},
  {Sick}, {Silbiger}, {Singanamalla}, {Singer}, {Sladen}, {Sooley},
  {Sornarajah}, {Streicher}, {Teuben}, {Thomas}, {Tremblay}, {Turner},
  {Terr{\'o}n}, {van Kerkwijk}, {de la Vega}, {Watkins}, {Weaver}, {Whitmore},
  {Woillez}, {Zabalza}, \& {Astropy Contributors}}]{astropy:2018}
{Astropy Collaboration}, {Price-Whelan}, A.~M., {Sip{\H{o}}cz}, B.~M., {et~al.}
  2018, \aj, 156, 123, \dodoi{10.3847/1538-3881/aabc4f}

\bibitem[{{Becker} {et~al.}(1995){Becker}, {White}, \& {Helfand}}]{becker1995}
{Becker}, R.~H., {White}, R.~L., \& {Helfand}, D.~J. 1995, \apj, 450, 559,
  \dodoi{10.1086/176166}

\bibitem[{{Begelman} {et~al.}(1979){Begelman}, {Rees}, \&
  {Blandford}}]{begelman1979}
{Begelman}, M.~C., {Rees}, M.~J., \& {Blandford}, R.~D. 1979, \nat, 279, 770,
  \dodoi{10.1038/279770a0}

\bibitem[{{Bhukta} {et~al.}(2021){Bhukta}, {Mondal}, \& {Pal}}]{bhukta2021}
{Bhukta}, N., {Mondal}, S.~K., \& {Pal}, S. 2021, arXiv e-prints,
  arXiv:2110.05484.
\newblock \doarXiv{2110.05484}

\bibitem[{{Blanton} {et~al.}(2001){Blanton}, {Gregg}, {Helfand}, {Becker}, \&
  {Leighly}}]{blanton2001}
{Blanton}, E.~L., {Gregg}, M.~D., {Helfand}, D.~J., {Becker}, R.~H., \&
  {Leighly}, K.~M. 2001, \aj, 121, 2915, \dodoi{10.1086/321074}

\bibitem[{{Blanton} {et~al.}(2000){Blanton}, {Gregg}, {Helfand}, {Becker}, \&
  {White}}]{blanton2000}
{Blanton}, E.~L., {Gregg}, M.~D., {Helfand}, D.~J., {Becker}, R.~H., \&
  {White}, R.~L. 2000, \apj, 531, 118, \dodoi{10.1086/308428}

\bibitem[{{Blanton} {et~al.}(2015){Blanton}, {Paterno-Mahler}, {Wing}, {Ashby},
  {Golden-Marx}, {Brodwin}, {Douglass}, {Randall}, \& {Clarke}}]{blanton2015}
{Blanton}, E.~L., {Paterno-Mahler}, R., {Wing}, J.~D., {et~al.} 2015, in
  Extragalactic Jets from Every Angle, ed. F.~{Massaro}, C.~C. {Cheung},
  E.~{Lopez}, \& A.~{Siemiginowska}, Vol. 313, 315--320,
  \dodoi{10.1017/S1743921315002410}

\bibitem[{{Burns} \& {Owen}(1980)}]{burnsowen1980}
{Burns}, J.~O., \& {Owen}, F.~N. 1980, \aj, 85, 204, \dodoi{10.1086/112663}

\bibitem[{{Cen} \& {Ostriker}(2006)}]{cenostriker2006}
{Cen}, R., \& {Ostriker}, J.~P. 2006, \apj, 650, 560, \dodoi{10.1086/506505}

\bibitem[{{Ching} {et~al.}(2017){Ching}, {Croom}, {Sadler}, {Robotham},
  {Brough}, {Baldry}, {Bland-Hawthorn}, {Colless}, {Driver}, {Holwerda},
  {Hopkins}, {Jarvis}, {Johnston}, {Kelvin}, {Liske}, {Loveday}, {Norberg},
  {Pracy}, {Steele}, {Thomas}, \& {Wang}}]{ching2017}
{Ching}, J.~H.~Y., {Croom}, S.~M., {Sadler}, E.~M., {et~al.} 2017, \mnras, 469,
  4584, \dodoi{10.1093/mnras/stx1173}

\bibitem[{{Condon} {et~al.}(1998){Condon}, {Cotton}, {Greisen}, {Yin},
  {Perley}, {Taylor}, \& {Broderick}}]{condon1998}
{Condon}, J.~J., {Cotton}, W.~D., {Greisen}, E.~W., {et~al.} 1998, \aj, 115,
  1693, \dodoi{10.1086/300337}

\bibitem[{{Croston} {et~al.}(2019){Croston}, {Hardcastle}, {Mingo}, {Best},
  {Sabater}, {Shimwell}, {Williams}, {Duncan}, {R{\"o}ttgering}, {Brienza},
  {G{\"u}rkan}, {Ineson}, {Miley}, {Morabito}, {O'Sullivan}, \&
  {Prandoni}}]{croston2019}
{Croston}, J.~H., {Hardcastle}, M.~J., {Mingo}, B., {et~al.} 2019, \aap, 622,
  A10, \dodoi{10.1051/0004-6361/201834019}

\bibitem[{{Dalal} {et~al.}(2021){Dalal}, {Strauss}, {Sunayama}, {Oguri}, {Lin},
  {Huang}, {Park}, \& {Takada}}]{dalal2021}
{Dalal}, R., {Strauss}, M.~A., {Sunayama}, T., {et~al.} 2021, \mnras, 507,
  4016, \dodoi{10.1093/mnras/stab2363}

\bibitem[{{De Lucia} \& {Blaizot}(2007)}]{delucia2007}
{De Lucia}, G., \& {Blaizot}, J. 2007, \mnras, 375, 2,
  \dodoi{10.1111/j.1365-2966.2006.11287.x}

\bibitem[{{de Vos} {et~al.}(2021){de Vos}, {Hatch}, {Merrifield}, \&
  {Mingo}}]{devos2021}
{de Vos}, K., {Hatch}, N.~A., {Merrifield}, M.~R., \& {Mingo}, B. 2021, \mnras,
  506, L55, \dodoi{10.1093/mnrasl/slab075}

\bibitem[{{Dey} {et~al.}(2019){Dey}, {Schlegel}, {Lang}, {Blum}, {Burleigh},
  {Fan}, {Findlay}, {Finkbeiner}, {Herrera}, {Juneau}, {Landriau}, {Levi},
  {McGreer}, {Meisner}, {Myers}, {Moustakas}, {Nugent}, {Patej}, {Schlafly},
  {Walker}, {Valdes}, {Weaver}, {Y{\`e}che}, {Zou}, {Zhou}, {Abareshi},
  {Abbott}, {Abolfathi}, {Aguilera}, {Alam}, {Allen}, {Alvarez}, {Annis},
  {Ansarinejad}, {Aubert}, {Beechert}, {Bell}, {BenZvi}, {Beutler}, {Bielby},
  {Bolton}, {Brice{\~n}o}, {Buckley-Geer}, {Butler}, {Calamida}, {Carlberg},
  {Carter}, {Casas}, {Castander}, {Choi}, {Comparat}, {Cukanovaite}, {Delubac},
  {DeVries}, {Dey}, {Dhungana}, {Dickinson}, {Ding}, {Donaldson}, {Duan},
  {Duckworth}, {Eftekharzadeh}, {Eisenstein}, {Etourneau}, {Fagrelius},
  {Farihi}, {Fitzpatrick}, {Font-Ribera}, {Fulmer}, {G{\"a}nsicke},
  {Gaztanaga}, {George}, {Gerdes}, {Gontcho}, {Gorgoni}, {Green}, {Guy},
  {Harmer}, {Hernandez}, {Honscheid}, {Huang}, {James}, {Jannuzi}, {Jiang},
  {Joyce}, {Karcher}, {Karkar}, {Kehoe}, {Kneib}, {Kueter-Young}, {Lan},
  {Lauer}, {Le Guillou}, {Le Van Suu}, {Lee}, {Lesser}, {Perreault Levasseur},
  {Li}, {Mann}, {Marshall}, {Mart{\'\i}nez-V{\'a}zquez}, {Martini}, {du Mas des
  Bourboux}, {McManus}, {Meier}, {M{\'e}nard}, {Metcalfe},
  {Mu{\~n}oz-Guti{\'e}rrez}, {Najita}, {Napier}, {Narayan}, {Newman}, {Nie},
  {Nord}, {Norman}, {Olsen}, {Paat}, {Palanque-Delabrouille}, {Peng},
  {Poppett}, {Poremba}, {Prakash}, {Rabinowitz}, {Raichoor}, {Rezaie},
  {Robertson}, {Roe}, {Ross}, {Ross}, {Rudnick}, {Safonova}, {Saha},
  {S{\'a}nchez}, {Savary}, {Schweiker}, {Scott}, {Seo}, {Shan}, {Silva},
  {Slepian}, {Soto}, {Sprayberry}, {Staten}, {Stillman}, {Stupak}, {Summers},
  {Sien Tie}, {Tirado}, {Vargas-Maga{\~n}a}, {Vivas}, {Wechsler}, {Williams},
  {Yang}, {Yang}, {Yapici}, {Zaritsky}, {Zenteno}, {Zhang}, {Zhang}, {Zhou}, \&
  {Zhou}}]{dey2019}
{Dey}, A., {Schlegel}, D.~J., {Lang}, D., {et~al.} 2019, \aj, 157, 168,
  \dodoi{10.3847/1538-3881/ab089d}

\bibitem[{{D'Onghia} {et~al.}(2005){D'Onghia}, {Sommer-Larsen}, {Romeo},
  {Burkert}, {Pedersen}, {Portinari}, \& {Rasmussen}}]{donghia2005}
{D'Onghia}, E., {Sommer-Larsen}, J., {Romeo}, A.~D., {et~al.} 2005, \apjl, 630,
  L109, \dodoi{10.1086/491651}

\bibitem[{{Duarte} \& {Mamon}(2014)}]{duartemamon2014}
{Duarte}, M., \& {Mamon}, G.~A. 2014, \mnras, 440, 1763,
  \dodoi{10.1093/mnras/stu378}

\bibitem[{{Edwards} {et~al.}(2010){Edwards}, {Fadda}, \&
  {Frayer}}]{edwards2010}
{Edwards}, L. O.~V., {Fadda}, D., \& {Frayer}, D.~T. 2010, \apjl, 724, L143,
  \dodoi{10.1088/2041-8205/724/2/L143}

\bibitem[{{Ekers}(1978)}]{ekers1978}
{Ekers}, R.~D. 1978, \aap, 69, 253

\bibitem[{{Falceta-Gon{\c{c}}alves} {et~al.}(2010){Falceta-Gon{\c{c}}alves},
  {Caproni}, {Abraham}, {Teixeira}, \& {de Gouveia Dal
  Pino}}]{falcetagoncalves2010}
{Falceta-Gon{\c{c}}alves}, D., {Caproni}, A., {Abraham}, Z., {Teixeira}, D.~M.,
  \& {de Gouveia Dal Pino}, E.~M. 2010, \apjl, 713, L74,
  \dodoi{10.1088/2041-8205/713/1/L74}

\bibitem[{{Freeland} {et~al.}(2008){Freeland}, {Cardoso}, \&
  {Wilcots}}]{freeland2008}
{Freeland}, E., {Cardoso}, R.~F., \& {Wilcots}, E. 2008, \apj, 685, 858,
  \dodoi{10.1086/591443}

\bibitem[{{Freeland} \& {Wilcots}(2011)}]{freeland2011}
{Freeland}, E., \& {Wilcots}, E. 2011, \apj, 738, 145,
  \dodoi{10.1088/0004-637X/738/2/145}

\bibitem[{{Garon} {et~al.}(2019){Garon}, {Rudnick}, {Wong}, {Jones}, {Kim},
  {Andernach}, {Shabala}, {Kapi{\'n}ska}, {Norris}, {de Gasperin}, {Tate}, \&
  {Tang}}]{garon2019}
{Garon}, A.~F., {Rudnick}, L., {Wong}, O.~I., {et~al.} 2019, \aj, 157, 126,
  \dodoi{10.3847/1538-3881/aaff62}

\bibitem[{{Golden-Marx} {et~al.}(2021){Golden-Marx}, {Blanton},
  {Paterno-Mahler}, {Brodwin}, {Ashby}, {Moravec}, {Shen}, {Lemaux}, {Lubin},
  {Gal}, \& {Tomczak}}]{goldenmarx2021}
{Golden-Marx}, E., {Blanton}, E.~L., {Paterno-Mahler}, R., {et~al.} 2021, \apj,
  907, 65, \dodoi{10.3847/1538-4357/abcd96}

\bibitem[{{Gozaliasl} {et~al.}(2019){Gozaliasl}, {Finoguenov}, {Tanaka},
  {Dolag}, {Montanari}, {Kirkpatrick}, {Vardoulaki}, {Khosroshahi}, {Salvato},
  {Laigle}, {McCracken}, {Ilbert}, {Cappelluti}, {Daddi}, {Hasinger}, {Capak},
  {Scoville}, {Toft}, {Civano}, {Griffiths}, {Balogh}, {Li}, {Ahoranta}, {Mei},
  {Iovino}, {Henriques}, \& {Erfanianfar}}]{gozaliasl2019}
{Gozaliasl}, G., {Finoguenov}, A., {Tanaka}, M., {et~al.} 2019, \mnras, 483,
  3545, \dodoi{10.1093/mnras/sty3203}

\bibitem[{{Heckman} \& {Best}(2014)}]{heckman2014}
{Heckman}, T.~M., \& {Best}, P.~N. 2014, \araa, 52, 589,
  \dodoi{10.1146/annurev-astro-081913-035722}

\bibitem[{{Hinshaw} {et~al.}(2013){Hinshaw}, {Larson}, {Komatsu}, {Spergel},
  {Bennett}, {Dunkley}, {Nolta}, {Halpern}, {Hill}, {Odegard}, {Page}, {Smith},
  {Weiland}, {Gold}, {Jarosik}, {Kogut}, {Limon}, {Meyer}, {Tucker}, {Wollack},
  \& {Wright}}]{hinshaw2013}
{Hinshaw}, G., {Larson}, D., {Komatsu}, E., {et~al.} 2013, \apjs, 208, 19,
  \dodoi{10.1088/0067-0049/208/2/19}

\bibitem[{{Huchra} \& {Geller}(1982)}]{huchrageller1982}
{Huchra}, J.~P., \& {Geller}, M.~J. 1982, \apj, 257, 423,
  \dodoi{10.1086/160000}

\bibitem[{{Jian} {et~al.}(2014){Jian}, {Lin}, {Chiueh}, {Lin}, {Liu}, {Merson},
  {Baugh}, {Huang}, {Chen}, {Foucaud}, {Murphy}, {Cole}, {Burgett}, \&
  {Kaiser}}]{jian2014}
{Jian}, H.-Y., {Lin}, L., {Chiueh}, T., {et~al.} 2014, \apj, 788, 109,
  \dodoi{10.1088/0004-637X/788/2/109}

\bibitem[{{Jones} \& {Owen}(1979)}]{jonesowen1979}
{Jones}, T.~W., \& {Owen}, F.~N. 1979, \apj, 234, 818, \dodoi{10.1086/157561}

\bibitem[{{Knowles} {et~al.}(2022){Knowles}, {Cotton}, {Rudnick}, {Camilo},
  {Goedhart}, {Deane}, {Ramatsoku}, {Bietenholz}, {Br{\"u}ggen}, {Button},
  {Chen}, {Chibueze}, {Clarke}, {de Gasperin}, {Ianjamasimanana}, {J{\'o}zsa},
  {Hilton}, {Kesebonye}, {Kolokythas}, {Kraan-Korteweg}, {Lawrie}, {Lochner},
  {Loubser}, {Marchegiani}, {Mhlahlo}, {Moodley}, {Murphy}, {Namumba},
  {Oozeer}, {Parekh}, {Pillay}, {Passmoor}, {Ramaila}, {Ranchod},
  {Retana-Montenegro}, {Sebokolodi}, {Sikhosana}, {Smirnov}, {Thorat},
  {Venturi}, {Abbott}, {Adam}, {Adams}, {Aldera}, {Bauermeister}, {Bennett},
  {Bode}, {Botha}, {Botha}, {Brederode}, {Buchner}, {Burger}, {Cheetham}, {de
  Villiers}, {Dikgale-Mahlakoana}, {du Toit}, {Esterhuyse}, {Fadana},
  {Fanaroff}, {Fataar}, {Foley}, {Fourie}, {Frank}, {Gamatham}, {Gatsi},
  {Geyer}, {Gouws}, {Gumede}, {Heywood}, {Hlakola}, {Hokwana}, {Hoosen},
  {Horn}, {Horrell}, {Hugo}, {Isaacson}, {Jonas}, {Jordaan}, {Joubert},
  {Julie}, {Kapp}, {Kasper}, {Kenyon}, {Kotz{\'e}}, {Kotze}, {Kriek}, {Kriel},
  {Krishnan}, {Kusel}, {Legodi}, {Lehmensiek}, {Liebenberg}, {Lord}, {Lunsky},
  {Madisa}, {Magnus}, {Main}, {Makhaba}, {Makhathini}, {Malan}, {Manley},
  {Marais}, {Maree}, {Martens}, {Mauch}, {McAlpine}, {Merry}, {Millenaar},
  {Mokone}, {Monama}, {Mphego}, {New}, {Ngcebetsha}, {Ngoasheng}, {Ockards},
  {Otto}, {Patel}, {Peens-Hough}, {Perkins}, {Ramanujam}, {Ramudzuli},
  {Ratcliffe}, {Renil}, {Robyntjies}, {Rust}, {Salie}, {Sambu}, {Schollar},
  {Schwardt}, {Schwartz}, {Serylak}, {Siebrits}, {Sirothia}, {Slabber},
  {Sofeya}, {Taljaard}, {Tasse}, {Tiplady}, {Toruvanda}, {Twum}, {van Balla},
  {van der Byl}, {van der Merwe}, {van Dyk}, {Van Tonder}, {Van Wyk}, {Venter},
  {Venter}, {Welz}, {Williams}, \& {Xaia}}]{knowles2022}
{Knowles}, K., {Cotton}, W.~D., {Rudnick}, L., {et~al.} 2022, \aap, 657, A56,
  \dodoi{10.1051/0004-6361/202141488}

\bibitem[{{Kozie{\l}-Wierzbowska} {et~al.}(2020){Kozie{\l}-Wierzbowska},
  {Goyal}, \& {{\.Z}ywucka}}]{kozielwierzbowska2020}
{Kozie{\l}-Wierzbowska}, D., {Goyal}, A., \& {{\.Z}ywucka}, N. 2020, \apjs,
  247, 53, \dodoi{10.3847/1538-4365/ab63d3}

\bibitem[{{Lacy} {et~al.}(2020){Lacy}, {Baum}, {Chandler}, {Chatterjee},
  {Clarke}, {Deustua}, {English}, {Farnes}, {Gaensler}, {Gugliucci},
  {Hallinan}, {Kent}, {Kimball}, {Law}, {Lazio}, {Marvil}, {Mao}, {Medlin},
  {Mooley}, {Murphy}, {Myers}, {Osten}, {Richards}, {Rosolowsky}, {Rudnick},
  {Schinzel}, {Sivakoff}, {Sjouwerman}, {Taylor}, {White}, {Wrobel},
  {Andernach}, {Beasley}, {Berger}, {Bhatnager}, {Birkinshaw}, {Bower},
  {Brandt}, {Brown}, {Burke-Spolaor}, {Butler}, {Comerford}, {Demorest}, {Fu},
  {Giacintucci}, {Golap}, {G{\"u}th}, {Hales}, {Hiriart}, {Hodge}, {Horesh},
  {Ivezi{\'c}}, {Jarvis}, {Kamble}, {Kassim}, {Liu}, {Loinard}, {Lyons},
  {Masters}, {Mezcua}, {Moellenbrock}, {Mroczkowski}, {Nyland}, {O'Dea},
  {O'Sullivan}, {Peters}, {Radford}, {Rao}, {Robnett}, {Salcido}, {Shen},
  {Sobotka}, {Witz}, {Vaccari}, {van Weeren}, {Vargas}, {Williams}, \&
  {Yoon}}]{lacy2020}
{Lacy}, M., {Baum}, S.~A., {Chandler}, C.~J., {et~al.} 2020, \pasp, 132,
  035001, \dodoi{10.1088/1538-3873/ab63eb}

\bibitem[{{Lang}(2014)}]{lang2014}
{Lang}, D. 2014, \aj, 147, 108, \dodoi{10.1088/0004-6256/147/5/108}

\bibitem[{{Lim} {et~al.}(2017){Lim}, {Mo}, {Wang}, \& {Yang}}]{lim2017}
{Lim}, S., {Mo}, H., {Wang}, H., \& {Yang}, X. 2017, arXiv e-prints.
\newblock \doarXiv{1712.08619}

\bibitem[{{Liu} {et~al.}(2008){Liu}, {Hsieh}, {Ho}, {Lin}, \& {Yan}}]{liu2008}
{Liu}, H.~B., {Hsieh}, B.~C., {Ho}, P. T.~P., {Lin}, L., \& {Yan}, R. 2008,
  \apj, 681, 1046, \dodoi{10.1086/588183}

\bibitem[{{Lopes} {et~al.}(2018){Lopes}, {Trevisan}, {Lagan{\'a}}, {Durret},
  {Ribeiro}, \& {Rembold}}]{lopes2018}
{Lopes}, P. A.~A., {Trevisan}, M., {Lagan{\'a}}, T.~F., {et~al.} 2018, \mnras,
  478, 5473, \dodoi{10.1093/mnras/sty1374}

\bibitem[{{Mingo} {et~al.}(2019){Mingo}, {Croston}, {Hardcastle}, {Best},
  {Duncan}, {Morganti}, {Rottgering}, {Sabater}, {Shimwell}, {Williams},
  {Brienza}, {Gurkan}, {Mahatma}, {Morabito}, {Prandoni}, {Bondi}, {Ineson}, \&
  {Mooney}}]{mingo2019}
{Mingo}, B., {Croston}, J.~H., {Hardcastle}, M.~J., {et~al.} 2019, \mnras, 488,
  2701, \dodoi{10.1093/mnras/stz1901}

\bibitem[{{Moravec} {et~al.}(2020){Moravec}, {Gonzalez}, {Stern}, {Clarke},
  {Brodwin}, {Decker}, {Eisenhardt}, {Mo}, {Pope}, {Stanford}, \&
  {Wylezalek}}]{moravec2020}
{Moravec}, E., {Gonzalez}, A.~H., {Stern}, D., {et~al.} 2020, \apj, 888, 74,
  \dodoi{10.3847/1538-4357/ab5af0}

\bibitem[{{Morsony} {et~al.}(2013){Morsony}, {Miller}, {Heinz}, {Freeland},
  {Wilcots}, {Br{\"u}ggen}, \& {Ruszkowski}}]{morsony2013}
{Morsony}, B.~J., {Miller}, J.~J., {Heinz}, S., {et~al.} 2013, \mnras, 431,
  781, \dodoi{10.1093/mnras/stt210}

\bibitem[{{Nelson} {et~al.}(2019){Nelson}, {Springel}, {Pillepich},
  {Rodriguez-Gomez}, {Torrey}, {Genel}, {Vogelsberger}, {Pakmor}, {Marinacci},
  {Weinberger}, {Kelley}, {Lovell}, {Diemer}, \& {Hernquist}}]{nelson2019}
{Nelson}, D., {Springel}, V., {Pillepich}, A., {et~al.} 2019, Computational
  Astrophysics and Cosmology, 6, 2, \dodoi{10.1186/s40668-019-0028-x}

\bibitem[{{Oogi} {et~al.}(2016){Oogi}, {Habe}, \& {Ishiyama}}]{oogi2016}
{Oogi}, T., {Habe}, A., \& {Ishiyama}, T. 2016, \mnras, 456, 300,
  \dodoi{10.1093/mnras/stv2581}

\bibitem[{{Owen} \& {Rudnick}(1976)}]{owenrudnick1976}
{Owen}, F.~N., \& {Rudnick}, L. 1976, \apjl, 205, L1, \dodoi{10.1086/182077}

\bibitem[{{Pal} \& {Kumari}(2021)}]{pal2021}
{Pal}, S., \& {Kumari}, S. 2021, arXiv e-prints, arXiv:2103.15199.
\newblock \doarXiv{2103.15199}

\bibitem[{{Pasini} {et~al.}(2021){Pasini}, {Finoguenov}, {Br{\"u}ggen},
  {Gaspari}, {de Gasperin}, \& {Gozaliasl}}]{pasini2021}
{Pasini}, T., {Finoguenov}, A., {Br{\"u}ggen}, M., {et~al.} 2021, \mnras,
  \dodoi{10.1093/mnras/stab1451}

\bibitem[{{Proctor}(2006)}]{proctor2006}
{Proctor}, D.~D. 2006, \apjs, 165, 95, \dodoi{10.1086/504801}

\bibitem[{{Riggs} {et~al.}(2021){Riggs}, {Barbhuiyan}, {Loveday}, {Brough},
  {Holwerda}, {Hopkins}, \& {Phillipps}}]{riggs2021}
{Riggs}, S.~D., {Barbhuiyan}, R.~W.~Y.~M., {Loveday}, J., {et~al.} 2021,
  \mnras, 506, 21, \dodoi{10.1093/mnras/stab1697}

\bibitem[{{Roberts} {et~al.}(2018){Roberts}, {Saripalli}, {Wang},
  {Sathyanarayana Rao}, {Subrahmanyan}, {KleinStern}, {Morii-Sciolla}, \&
  {Simpson}}]{roberts2018}
{Roberts}, D.~H., {Saripalli}, L., {Wang}, K.~X., {et~al.} 2018, \apj, 852, 47,
  \dodoi{10.3847/1538-4357/aa9c49}

\bibitem[{{Sakelliou} \& {Merrifield}(2000)}]{sakelliou2000}
{Sakelliou}, I., \& {Merrifield}, M.~R. 2000, \mnras, 311, 649,
  \dodoi{10.1046/j.1365-8711.2000.03079.x}

\bibitem[{{Sanderson} {et~al.}(2013){Sanderson}, {O'Sullivan}, {Ponman},
  {Gonzalez}, {Sivanandam}, {Zabludoff}, \& {Zaritsky}}]{sanderson2013}
{Sanderson}, A.~J.~R., {O'Sullivan}, E., {Ponman}, T.~J., {et~al.} 2013,
  \mnras, 429, 3288, \dodoi{10.1093/mnras/sts586}

\bibitem[{{Shen} {et~al.}(2014){Shen}, {Yang}, {Mo}, {van den Bosch}, \&
  {More}}]{shen2014}
{Shen}, S., {Yang}, X., {Mo}, H., {van den Bosch}, F., \& {More}, S. 2014,
  \apj, 782, 23, \dodoi{10.1088/0004-637X/782/1/23}

\bibitem[{{Shimwell} {et~al.}(2017){Shimwell}, {R{\"o}ttgering}, {Best},
  {Williams}, {Dijkema}, {de Gasperin}, {Hardcastle}, {Heald}, {Hoang},
  {Horneffer}, {Intema}, {Mahony}, {Mandal}, {Mechev}, {Morabito}, {Oonk},
  {Rafferty}, {Retana-Montenegro}, {Sabater}, {Tasse}, {van Weeren},
  {Br{\"u}ggen}, {Brunetti}, {Chy{\.z}y}, {Conway}, {Haverkorn}, {Jackson},
  {Jarvis}, {McKean}, {Miley}, {Morganti}, {White}, {Wise}, {van Bemmel},
  {Beck}, {Brienza}, {Bonafede}, {Calistro Rivera}, {Cassano}, {Clarke},
  {Cseh}, {Deller}, {Drabent}, {van Driel}, {Engels}, {Falcke}, {Ferrari},
  {Fr{\"o}hlich}, {Garrett}, {Harwood}, {Heesen}, {Hoeft}, {Horellou},
  {Israel}, {Kapi{\'n}ska}, {Kunert-Bajraszewska}, {McKay}, {Mohan},
  {Orr{\'u}}, {Pizzo}, {Prandoni}, {Schwarz}, {Shulevski}, {Sipior}, {Smith},
  {Sridhar}, {Steinmetz}, {Stroe}, {Varenius}, {van der Werf}, {Zensus}, \&
  {Zwart}}]{shimwell2017}
{Shimwell}, T.~W., {R{\"o}ttgering}, H.~J.~A., {Best}, P.~N., {et~al.} 2017,
  \aap, 598, A104, \dodoi{10.1051/0004-6361/201629313}

\bibitem[{{Silverstein} {et~al.}(2018){Silverstein}, {Anderson}, \&
  {Bregman}}]{silverstein2018}
{Silverstein}, E.~M., {Anderson}, M.~E., \& {Bregman}, J.~N. 2018, \aj, 155,
  14, \dodoi{10.3847/1538-3881/aa9d2e}

\bibitem[{{Solanes} {et~al.}(2016){Solanes}, {Perea}, {Darriba},
  {Garc{\'\i}a-G{\'o}mez}, {Bosma}, \& {Athanassoula}}]{solanes2016}
{Solanes}, J.~M., {Perea}, J.~D., {Darriba}, L., {et~al.} 2016, \mnras, 461,
  321, \dodoi{10.1093/mnras/stw1278}

\bibitem[{{Tempel} {et~al.}(2014){Tempel}, {Tamm}, {Gramann}, {Tuvikene},
  {Liivam{\"a}gi}, {Suhhonenko}, {Kipper}, {Einasto}, \& {Saar}}]{tempel2014}
{Tempel}, E., {Tamm}, A., {Gramann}, M., {et~al.} 2014, \aap, 566, A1,
  \dodoi{10.1051/0004-6361/201423585}

\bibitem[{{Wing} \& {Blanton}(2011)}]{wingblanton2011}
{Wing}, J.~D., \& {Blanton}, E.~L. 2011, \aj, 141, 88,
  \dodoi{10.1088/0004-6256/141/3/88}

\bibitem[{{Wing} \& {Blanton}(2013)}]{wingblanton2013}
---. 2013, \apj, 767, 102, \dodoi{10.1088/0004-637X/767/2/102}

\bibitem[{{Worpel} {et~al.}(2013){Worpel}, {Brown}, {Jones}, {Floyd}, \&
  {Beutler}}]{worpel2013}
{Worpel}, H., {Brown}, M. J.~I., {Jones}, D.~H., {Floyd}, D. J.~E., \&
  {Beutler}, F. 2013, \apj, 772, 64, \dodoi{10.1088/0004-637X/772/1/64}

\bibitem[{{Yang} {et~al.}(2008){Yang}, {Mo}, \& {van den Bosch}}]{yang2008}
{Yang}, X., {Mo}, H.~J., \& {van den Bosch}, F.~C. 2008, \apj, 676, 248,
  \dodoi{10.1086/528954}

\bibitem[{{Zirbel}(1997)}]{zirbel1997}
{Zirbel}, E.~L. 1997, \apj, 476, 489, \dodoi{10.1086/303626}

\bibitem[{{Zou} {et~al.}(2019){Zou}, {Gao}, {Zhou}, \& {Kong}}]{zou2019}
{Zou}, H., {Gao}, J., {Zhou}, X., \& {Kong}, X. 2019, \apjs, 242, 8,
  \dodoi{10.3847/1538-4365/ab1847}

\end{thebibliography}
\bibliographystyle{aasjournal}

\end{document}